\documentclass[sigconf,screen]{acmart}

\usepackage{todonotes}
\usepackage{xspace}
\usepackage[capitalise]{cleveref}
\usepackage{balance}
\usepackage{booktabs}
\usepackage{tabularx}
\usepackage{subcaption}

\AtBeginDocument{%
	}

\copyrightyear{2025}
\acmYear{2025}
\setcopyright{acmlicensed}\acmConference[FSE Companion '25]{33rd ACM International Conference on the Foundations of Software Engineering}{June 23--28, 2025}{Trondheim, Norway}
\acmBooktitle{33rd ACM International Conference on the Foundations of Software Engineering (FSE Companion '25), June 23--28, 2025, Trondheim, Norway}
\acmDOI{10.1145/3696630.3727231}
\acmISBN{979-8-4007-1276-0/2025/06}

\begin{document}
	
	\title[Sojourner under Sabotage: A Serious Testing and Debugging Game]{Sojourner under Sabotage: \\A Serious Testing and Debugging Game}
	
	\author{Philipp Straubinger}
	\affiliation{%
		\institution{University of Passau}
		\country{Germany}}
	
	\author{Tim Greller}
	\affiliation{%
		\institution{University of Passau}
		\country{Germany}}
	
	\author{Gordon Fraser}
	\affiliation{%
		\institution{University of Passau}
		\country{Germany}}
	
	\renewcommand{\shortauthors}{Straubinger et al.}
	\newcommand{\toolname}{\emph{Sojourner under Sabotage}\xspace}
	\newcommand{\passau}{University of Passau\xspace}
	
	\newcommand{\summary}[2]{%
		\vspace{-0.2cm}%
		\begin{center}%
			\colorbox{gray!20}{%
				\parbox{\linewidth}{%
					\textbf{\textsf{Summary (\textit{#1})}:}~%
					#2%
				}%
			}%
		\end{center}%
	}
	
	\begin{abstract}
		Teaching software testing and debugging is a critical yet challenging task in computer science education, often hindered by low student engagement and the perceived monotony of these activities. \toolname, a browser-based serious game, reimagines this learning experience by blending education with an immersive and interactive storyline. Players take on the role of a spaceship crew member, using unit testing and debugging techniques to identify and repair sabotaged components across seven progressively challenging levels. A study with 79 students demonstrates that the game is a powerful tool for enhancing motivation, engagement, and skill development. These findings underscore the transformative potential of serious games in making essential software engineering practices accessible and enjoyable.
	\end{abstract}

	
	
	\maketitle
	
	\section{Introduction}
	\begin{figure}[t]
	\centering
	\includegraphics[width=\linewidth]{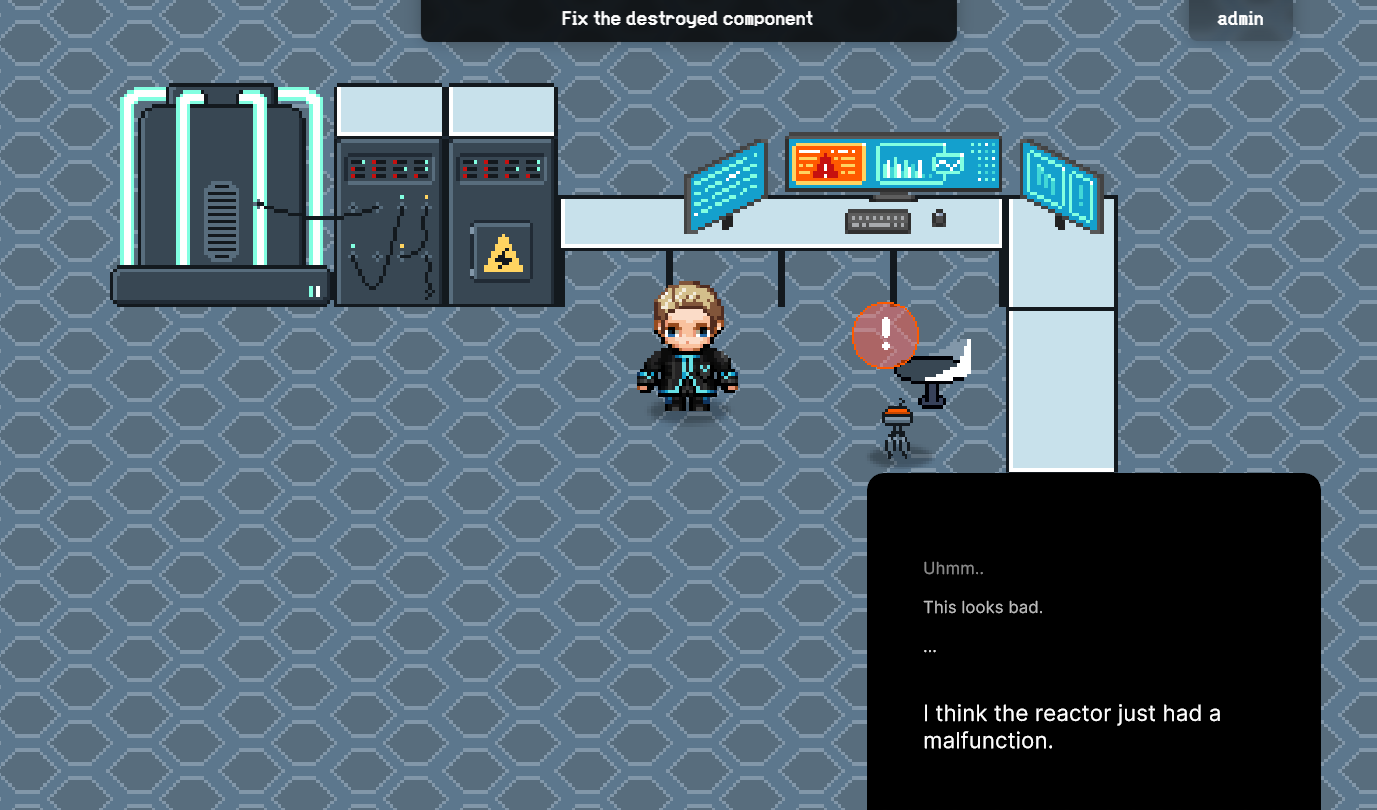}
	\caption{\toolname during an interaction between the player and the accompanying robot, which is reporting a sabotaged component}
	\label{fig:player}
\end{figure}

\begin{figure}[t]
	\centering
	\includegraphics[width=\linewidth]{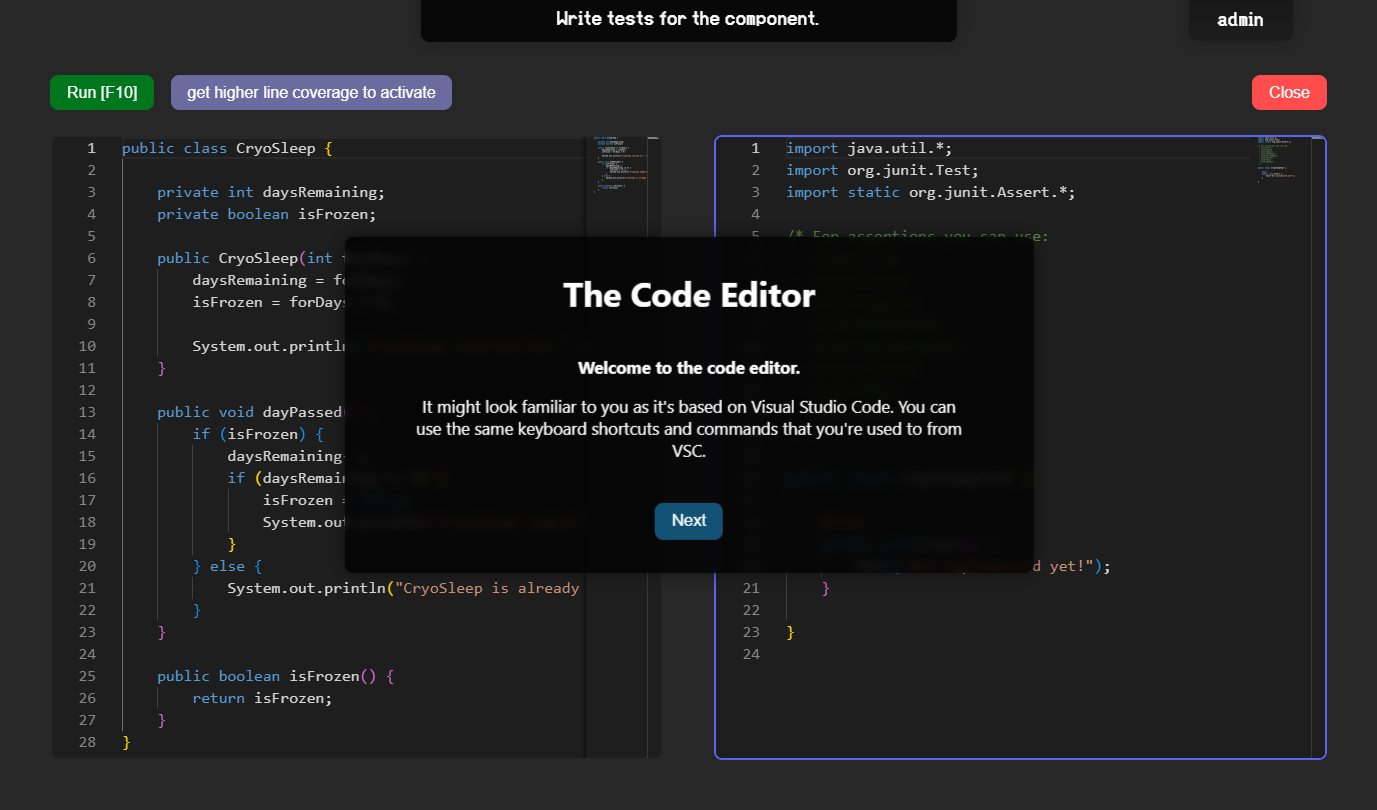}
	\caption{Code editor integrated into the game for writing tests (shown in picture), but also for debugging and fixing}
	\label{fig:editor}
\end{figure}

Teaching software testing and debugging poses significant challenges in computer science education~\cite{park2025exploring,DBLP:conf/issre/StraubingerF23,garousi2020software}. Students often perceive these tasks as tedious and less rewarding than programming, hindering their engagement and motivation~\cite{park2025exploring,garousi2020software}. However, mastering these skills is critical for developing reliable software, as they form the backbone of modern software engineering practices~\cite{7814898,DBLP:journals/csedu/McCauleyFLMSTZ08}. Serious games, which combine educational objectives with game-based mechanics, offer a promising approach to overcoming these challenges~\cite{DBLP:conf/cist/YamoulOMR23,DBLP:journals/ce/ConnollyBMHB12}. By gamifying testing and debugging, educators can foster intrinsic motivation, engage students with hands-on activities, and help them develop essential skills in a more enjoyable and interactive way~\cite{DBLP:journals/ieee-rita/QuinteroA23,DBLP:journals/jss/BlancoTCCRT23}.

In this paper, we introduce \toolname, a browser-based serious game designed to teach and reinforce testing and debugging skills entertainingly. \toolname is a puzzle game in which players assume the role of a spaceship crew member and navigate the spaceship in the style of a role-playing game, shown in \cref{fig:player}. As the storyline unfolds, players have to identify and repair sabotaged components using unit testing and debugging techniques, in an IDE-like environment integrated into the game (\cref{fig:editor}). The game features seven levels of increasing difficulty, incorporating elements such as mutation testing, code tracing, and test coverage visualization. Players write tests, identify faults through their execution results, and fix bugs within a gamified environment that includes a storyline, a robot companion, and engaging mini-games.

To evaluate the effectiveness of \toolname, we conducted two controlled sessions with undergraduate students -- one with first-year and another with third-year students -- to capture a range of programming experience. By comparing these groups, we examined how students engaged with the game, their performance on testing and debugging tasks, and their overall perceptions of the experience, in order to assess the game's educational value.

In detail, the contributions of this work are as follows:
\begin{itemize}
	\item The development of \toolname, a novel serious game that integrates unit testing and debugging into an engaging educational context.
	\item The design and implementation of two studies with 79 participants with two different educational backgrounds.
	\item A comprehensive evaluation of the game's effectiveness, focusing on student engagement, skill development, and perceived difficulty.
	\item Insights into how \toolname can address the challenges of teaching testing and debugging in computer science education.  
\end{itemize}

Our findings show that students enjoyed the game, with over 80\% expressing satisfaction with its design and educational value. The game successfully motivated students to write tests and debug code, with more experienced participants achieving higher test coverage and progression through levels. However, differences in experience influenced task difficulty, with less experienced students finding debugging more challenging. These results highlight the potential of serious games to improve testing and debugging education while also identifying areas for refinement in game design.

	\section{Background}
	
\toolname is a serious game that aims to support education of two core
aspects of software engineering education: Testing and debugging.



\subsection{Software Testing Education}

If software testing is taught, then sometimes this is done as a standalone course, but more often it is included within general programming or software engineering classes. Ideally, even when dedicated courses exist, testing should be integrated into all programming courses and introduced early in a student's education~\cite{DBLP:journals/jss/GarousiRLA20}.  

A key challenge in teaching software testing, however, is overcoming the perception that writing tests is tedious, boring, and redundant compared to the subjectively more rewarding task of developing programs~\cite{DBLP:journals/jss/GarousiRLA20,DBLP:conf/issre/StraubingerF23}. Educators must therefore focus on inspiring and motivating students~\cite{DBLP:journals/jss/BlancoTCCRT23}. In particular, in an industrial setting, software testers are often driven by a desire for knowledge, variety, and creativity~\cite{DBLP:conf/esem/SantosMCSCS17}, which can be supported by raising initial motivation already during education.

\subsection{Debugging Education}

Effective debugging requires domain knowledge (understanding the programming language), system knowledge (program structure and interactions), procedural knowledge (using tools like IDEs), strategic knowledge (debugging strategies), and prior experience. Teaching these elements systematically is thought to improve students’ ability to identify and fix errors~\cite{DBLP:conf/ace/LiCDLT19}.

However, like testing, debugging is often underemphasized in computer science education, with minimal guidance on how to teach it effectively~\cite{DBLP:journals/csedu/McCauleyFLMSTZ08,DBLP:conf/wipsce/MichaeliR19,DBLP:conf/sigcse/MurphyLMSTZ08} despite its inclusion in the ACM/IEEE curriculum~\cite{DBLP:journals/inroads/RajK22}. Novice programmers often struggle with adapting to programming languages due to misconceptions, such as assuming the language will interpret their code as intended, a challenge exacerbated by differences between natural and programming languages~\cite{pea1986language,DBLP:journals/hhci/BonarS85}.
Common bugs include boundary errors, misplaced code, logical flaws, and calculation mistakes, often stemming from confusion or gaps in understanding. Addressing these issues in education can help students build better debugging skills~\cite{DBLP:journals/csedu/McCauleyFLMSTZ08}.  

\subsection{Gamification and Serious Games}

Gamification refers to the integration of game elements into non-game
tasks to boost motivation and
engagement~\cite{DBLP:conf/mindtrek/DeterdingDKN11,DBLP:books/sp/23/Cooper23}.
Although gamification is effective in increasing
engagement in software testing education, there can also be
drawbacks~\cite{DBLP:conf/hefa/TodaVI17,DBLP:conf/gamification/KappenN13}:
Poorly designed approaches, such as over-reliance on
point-badge-leaderboard systems, can lead to demotivation,
frustration, or disengagement, and negative effects often arise from
penalties, overly complex rules, or fear of academic consequences.
In order to counter these issues, it is important to prioritize and support intrinsic motivation---driven by interest,
achievement, and emotional satisfaction. Elements such as autonomy, competence, and relatedness
help sustain engagement and foster meaningful learning
experiences. Integrating values that resonate with students further
enhances commitment and skills
development~\cite{DBLP:conf/gamification/KappenN13}. One way to put
focus on some of these aspects is through the use of \emph{serious games}.

Serious games go beyond gamification in that they not only include game elements in regular tasks, but are full video games designed to develop skills, knowledge, and behavioral changes through gameplay, rather than focusing solely on entertainment~\cite{DBLP:journals/ce/ConnollyBMHB12,DBLP:books/sp/23/Cooper23}. Since the 2000s, serious games have become increasingly prominent in education, showing positive effects on student motivation, learning, and performance~\cite{DBLP:conf/cist/YamoulOMR23,DBLP:journals/ce/ConnollyBMHB12}. They help develop both subject-specific and general competencies, enhancing educational outcomes~\cite{DBLP:journals/ieee-rita/QuinteroA23,DBLP:journals/jss/BlancoTCCRT23}.
However, creating serious games is costly and complex, requiring teacher training and serving as a supplement rather than a replacement for traditional teaching methods~\cite{DBLP:conf/cist/YamoulOMR23}. 


	\section{Sojourner under Sabotage}
	\begin{figure}[t]
	\centering
	\includegraphics[width=\linewidth]{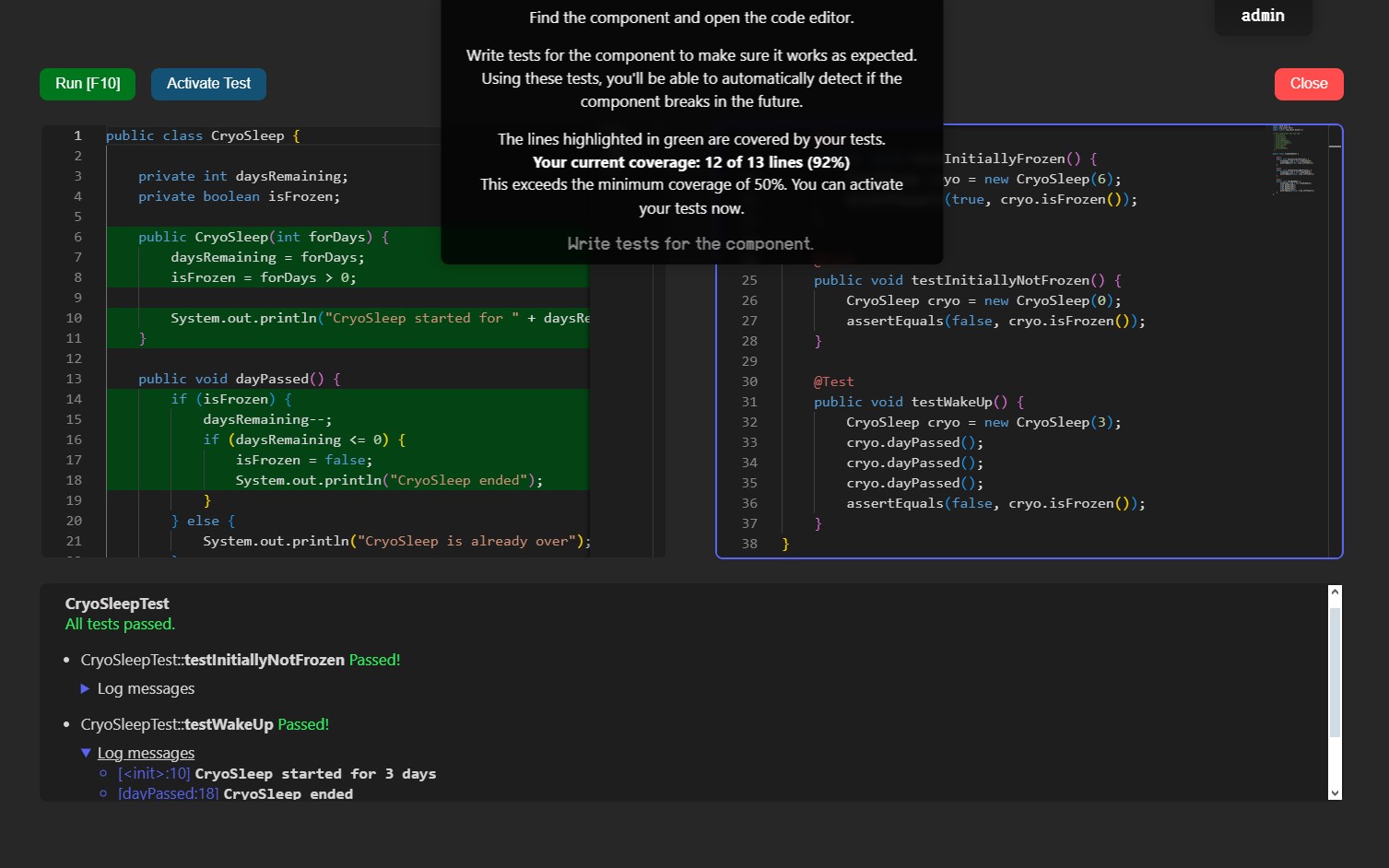}
	\caption{Code editor of \toolname after writing tests}
	\label{fig:finishedtests}
\end{figure}

\begin{figure}[t]
	\centering
	\includegraphics[width=\linewidth]{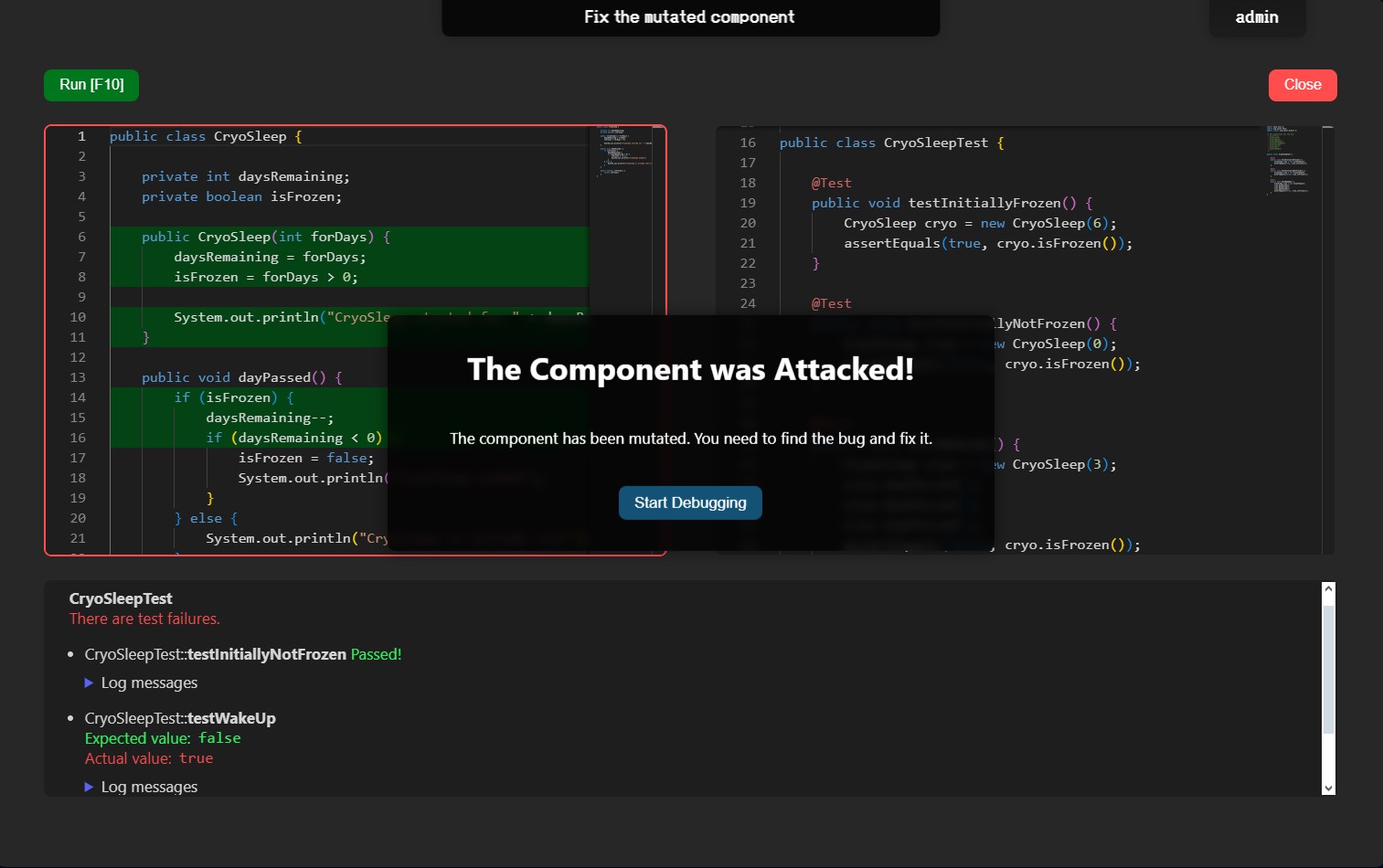}
	\caption{Code editor of \toolname while debugging}
	\label{fig:debug}
\end{figure}

In \toolname, the player takes on the role of a spaceship crew member who wakes up early due to a malfunction in their cryogenic chamber. With the help of a robot companion, the player must secure the spaceship by identifying and fixing sabotaged components. Unit tests are used to detect sabotages: when a component fails, an alarm is triggered, prompting the player to locate and debug the faulty code. Each of the seven levels involves repairing a unique component in a specific room by writing tests, detecting sabotage, and fixing the issue. Throughout the game, the player navigates the spaceship, interacts with objects, and is followed closely by the robot.

At the start of the game, the player wakes up in the cryogenic chamber, unaware of what has happened. They must first interact with the robot, which explains that the cryogenic capsule malfunctioned and that the ship needs to be secured by writing tests to detect further issues. The player’s first task is to find the room’s console, where they can open the editor and begin writing tests, as shown in \cref{fig:editor}.

\subsection{Writing tests}

Once the editor is opened (\cref{fig:editor}), a brief tutorial explains how it works. The editor has two text areas: one for the component’s source code (to test and later fix) and another for the player’s test code. The active editor is highlighted with a colored border. Controls above the editor let the player execute tests, activate these tests once enough coverage is achieved, or return to the spaceship view.

For example, in the first level, the player has to investigate the \texttt{CryoSleep} class, (shown in \cref{fig:editor} and successive screenshots) which represents a cryogenic sleep system that keeps track of the remaining days and whether the system is currently active. The constructor initializes the system by setting the number of remaining days and activating it if the duration is greater than zero. The \texttt{dayPassed()} method simulates the passage of one day, reducing the remaining days if the system is active. When the days reach zero, the system is deactivated. If the system is already inactive, it displays a message. The \texttt{isFrozen()} method checks and returns whether the cryogenic sleep system is still active.

The player has to write tests for this class to ensure its functionality. Running these tests displays results in a console panel, including logs and error messages, for example shown in \cref{fig:finishedtests} for the \texttt{CryoSleep} class of level 1. A pop-up window notifies the player when their tests first succeed and explains that at least 50\% coverage is required to activate the tests. Covered lines are highlighted in green, and once the threshold is met, the \textit{activate} button becomes available. Once the player presses this button, the tests actively check the component for bugs resulting from sabotage.

\subsection{Debugging the component}

Once the player activates their tests, they ensure the component is functioning correctly, allowing the player to explore the room and interact with objects or the robot. After some time, the component is sabotaged, mutating its code to introduce a bug. If the tests detect the sabotage, an alarm is triggered, and the robot alerts the player (shown in \cref{fig:player}). If the tests fail to catch the sabotage, the component is destroyed. In this case, the robot repairs the physical damage and adds a new test to the suite, pointing out the bug’s effects to keep the gameplay balanced and reduce frustration.

To debug the component, the player must return to the console. The editor now displays the mutated code, with any failed tests highlighted in red, as shown in \cref{fig:debug}. A brief introduction helps the player identify the bug. The editor also provides test results and detailed failure messages, supporting various debugging strategies such as analyzing failed assertions, tracing code statically, or using print statements. Logs are organized by test method, and visual code coverage indicates which parts of the code were executed, helping the player isolate the issue more effectively.

For example, in the \texttt{CryoSleep} class, the sabotage involves changing the \texttt{<=} operator to \texttt{<} in the \texttt{dayPassed()} method. This causes the system to wake only when the remaining days are strictly less than zero rather than zero or less. After fixing the bug such that all tests pass, hidden tests are run to ensure no new bugs are introduced. If hidden tests fail, they are added to the player’s test suite for further debugging. Once all tests pass, the component is repaired, and the level is complete.

\subsection{Minigames}

\begin{figure}[t]
	\centering
	\includegraphics[width=0.8\linewidth]{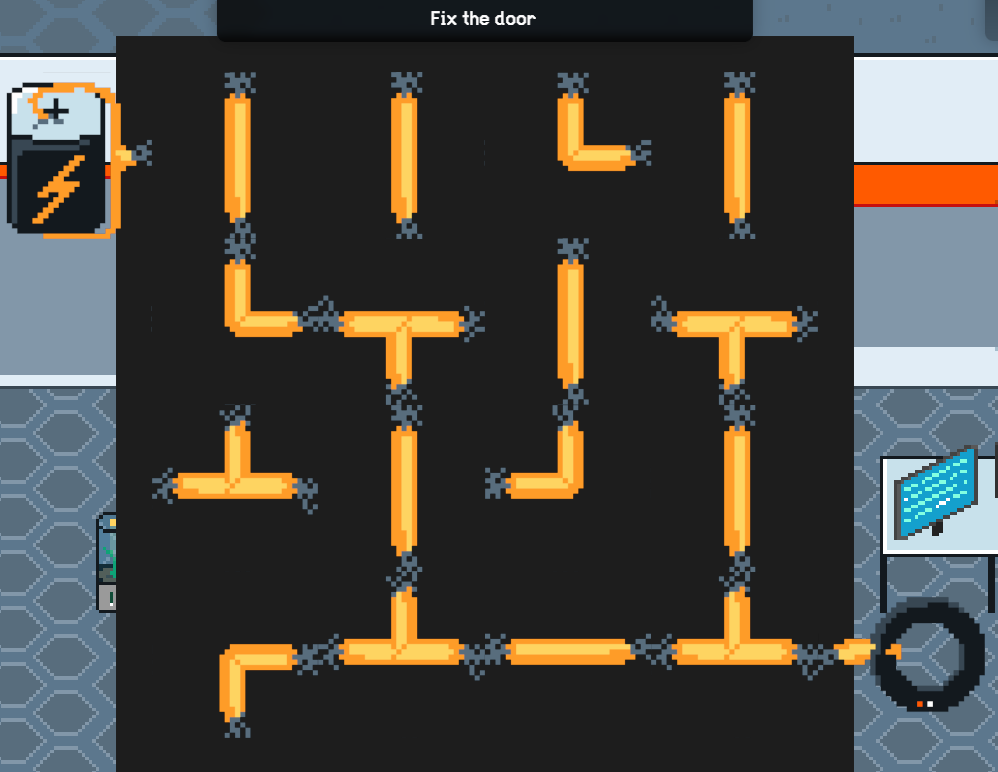}
	\caption{Minigame of \toolname to open doors between levels}
	\label{fig:minigame}
\end{figure}

After repairing a component, the player advances to the next level. To unlock the next room, they must complete a short minigame. This involves solving a puzzle where the player rotates wires to connect all parts of an electrical circuit between the power source and the door mechanism (\cref{fig:minigame}). These puzzles vary in size and complexity, adding variety to the gameplay and keeping players engaged. The minigames also provides a mental break from testing and debugging, keeping the experience fresh and motivating.

\subsection{Levels}

\begin{table*}[t!]
	\small
	\centering
	\caption{Levels in \toolname}
	\begin{tabularx}{\textwidth}{clXX}
		\toprule
		\textbf{Level} & \textbf{Name} & \textbf{Programming Concept} & \textbf{Bug Description} \\ \midrule
		1 & Cryo Chamber & Boundary conditions and state transitions. & Off-by-one error in checking the number of days remaining, causing the pod to wake up too late. \\ 
		2 & Engine Compartment & Floating-point calculations and JUnit assertions with deltas. & Spurious code wrapping a double argument in a Math.floor call, requiring floating-point testing. \\ 
		3 & Green House & Arrays, switch statements, and exception handling. & Missing break statement in a switch case, causing dead plants to be replanted automatically. \\ 
		4 & Kitchen & Map iteration and algorithmic execution order. & Misplaced code causing partial execution of a recipe due to ingredient removal in the loop. \\ 
		5 & Reactor & Lists and boundary conditions in data processing. & Logic error where the latest maximum temperature is not logged due to a malformed condition. \\ 
		6 & Defense System & Spatial calculations and mathematical operations. & Misplaced code swapping x and y values in calculations, leading to incorrect spatial adjustments. \\ 
		7 & Infirmary, RNA-Analyzer & Recursion and string manipulation. & Swapped arguments in a recursive call, causing incorrect substring matching. \\ \bottomrule
	\end{tabularx}
	\label{tab:levels}
\end{table*}

\begin{figure*}[t]
	\centering
	\includegraphics[width=\linewidth]{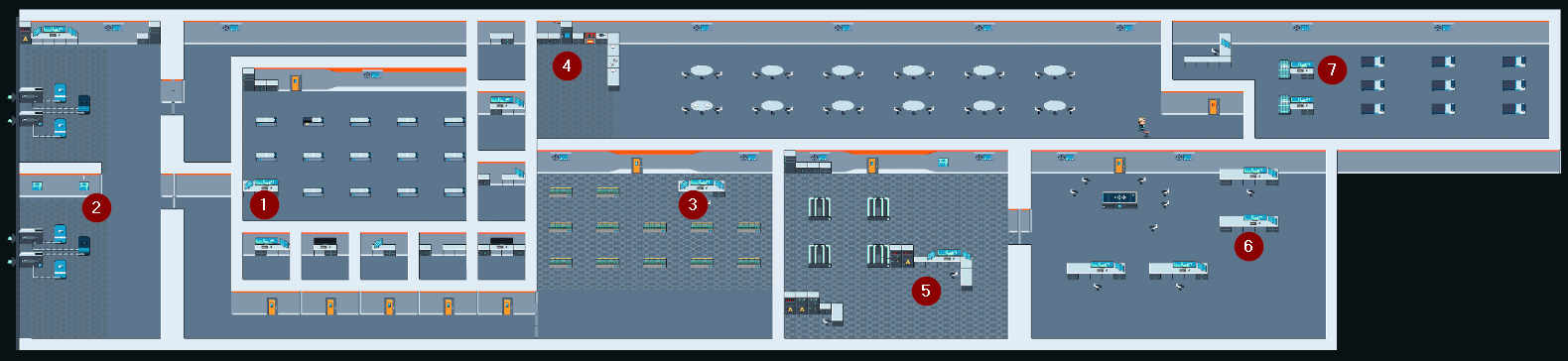}
	\caption{Complete map of \toolname including all levels}
	\label{fig:map}
\end{figure*}

The game features seven progressively challenging levels (\cref{tab:levels}), all in a single large map, with component locations marked on the map (see \cref{fig:map}). The source code for each component is designed to fit the spaceship context, be easy for students to understand, and be suitable for JUnit testing. Levels cover various programming concepts: the first is introductory, the second and sixth focus on floating-point math (with level six being more advanced), levels three to five address loops and Java data structures (arrays, lists, and maps), and level seven involves a recursive algorithm.

Each level contains one program anomaly issue to fix, restricted to semantic or logical errors at a single location (type 1 in Nayrolles and Hamou-Lhadj’s classification~\cite{DBLP:conf/icse/NayrollesH18}). These issues reflect common misconceptions or flaws in algorithmic thinking, avoiding syntax errors that would merely trigger a compiler failure~\cite{DBLP:conf/sigcse/HristovaMRM03}. Defects are categorized as missing, spurious, misplaced, or malformed code, aligning with the principles of unit testing and focusing on standalone classes rather than broader system interactions\cite{DBLP:journals/jss/CatolinoPZF19}.

\subsection{Implementation}

\toolname is a browser-based game, leveraging WebGL\footnote{\url{https://www.khronos.org/webgl/}} for compatibility across operating systems without requiring downloads or installations. The Unity engine,\footnote{\url{https://unity.com/}} using C\# and the OneJS library\footnote{\url{https://onejs.com/}} for TypeScript and UI creation with Preact,\footnote{\url{https://preactjs.com/}} is applied to build the game, which is then exported to WebGL. The backend, implemented in Java with Spring Boot,\footnote{\url{https://spring.io/projects/spring-boot}} handles game logic, APIs, and code execution. Code execution uses the Java Compiler API and is hosted on a Tomcat server.\footnote{\url{https://tomcat.apache.org/}}

The game features a 2D top-down perspective. Aiding gameplay, the robot companion follows the player using the A* pathfinding algorithm~\cite{cui2011based} combined with Reynolds' path-following technique~\cite{reynolds1999steering}. The Monaco Editor,\footnote{\url{https://github.com/microsoft/monaco-editor}} embedded for in-game coding, provides advanced features like syntax highlighting, autocompletion, and error visualization, ensuring an efficient and familiar environment for users.

Game mechanics focus on the dynamic execution of user-written Java code. During gameplay, tests are executed to detect sabotage and verify repairs. The execution service manages these processes, using JUnitCore\footnote{\url{https://junit.org/junit4/javadoc/latest/org/junit/runner/JUnitCore.html}} for test execution, including capturing results, stack traces, and sandbox violations. To prevent infinite loops, a thread-timer mechanism terminates long-running code, ensuring seamless gameplay.

	\section{Experiment Setup}
	To evaluate \toolname, we conducted a controlled experiment aiming to answer the following research questions:

\begin{itemize}
	\item \textbf{RQ 1}: How do students engage with \toolname?
	\item \textbf{RQ 2}: How do students perform in testing activities with \toolname?
	\item \textbf{RQ 3}: How do students perform in debugging activities with \toolname?
	\item \textbf{RQ 4}: How do students perceive \toolname?
\end{itemize}

\subsection{Experiment Procedure}

We conducted two sessions in May and November 2024, where students were invited to play \toolname.
Each participant was provided with a user account. 
At the start of each session, we gave students an introductory overview of \toolname, explaining how to navigate the game and interact with its components, including the robot and the editor. Students were allotted a total of 60 minutes to play the game during each session. While we encouraged active participation, we did not require them to complete specific tasks. As an incentive, they received a 5\% bonus on their grade for participating. No additional introduction to JUnit was provided, as this topic had been covered in a lecture held a few days earlier. During the session, students could ask questions and use the internet for assistance, such as looking up programming language documentation or JUnit references. However, they were not allowed to use any Large Language Models. After the sessions, students were asked to complete a survey that collected demographic information, and feedback on \toolname, and included a brief questionnaire (see \cref{fig:survey}).

\subsection{Participants} \label{sec:participants}

The participants in the first session were enrolled in a general first-year software engineering (SE) course at the \passau. A total of 45 students took part, with 27\% identifying as female and the rest as male. Their ages ranged from 18 to 38, with the majority being between 19 and 22 years. Over half of the participants studied computer science, while 14\% were enrolled in internet computing and 29\% in business computer science. Programming experience varied widely, with some participants having less than three months of experience and others reporting more than three years.

The second session's participants were part of a specialized software testing (ST) course taught during their third year at the \passau. This session included 34 students, of whom 15\% identified as female and the rest as male. Their ages ranged from 19 to 29, with most falling between 20 and 22 years. Approximately three-quarters of the participants studied computer science, 18\% internet computing, and 9\% educational studies. All participants had at least one year of programming experience, with two-thirds reporting more than three years. Additionally, roughly half stated they had more than one year of experience with software testing.

\subsection{Experiment Analysis}

Throughout the experiments we collected data on the actions performed by the players, the tests they wrote, and finally also their survey responses. When comparing  between the two groups SE and ST,
%
we use the exact Wilcoxon-Mann-Whitney test~\cite{10.1214/aoms/1177730491} to calculate the $p$-values with $\alpha = 0.05$. 

\paragraph{RQ 1: How do students engage with \toolname?}

To answer this research question, we investigate (1) the overall time spent on testing, debugging, and other game activities, (2) the time spent on testing and debugging per level, and (3) the execution results obtained during gameplay.

\paragraph{RQ 2: How do students perform in testing activities with \toolname?}

To answer this research question, we inspect the (1) line coverage, (2) mutation score, (3) the number of tests written, (4) the presence of test smells, and (5) the effectiveness of their test suites, measured by how often the seeded bugs were detected. Coverage is measured in real-time during gameplay using bytecode instrumentation, while mutation scores are calculated using PIT,\footnote{\url{https://pitest.org/}} and test smells are identified with the Test Smell Detector.\footnote{\url{https://testsmells.org/}}

\paragraph{RQ 3: How do students perform in debugging activities with \toolname?}

To answer this research question, we inspect (1) the time spent on modifications per level, (2) the number of newly introduced bugs, and (3)~the usage of print statements during debugging.

\paragraph{RQ 4: How do students perceive \toolname?}

This research question is answered using the responses to the exit survey.

\subsection{Threats to Validity}

The \textit{internal validity} of our study might be affected by the differences in prior knowledge and programming experience among participants. Although the two groups were selected from different courses—one focused on software engineering and the other on software testing—the variation in their exposure to relevant material and practical experience could influence the results. Additionally, the time gap between the two sessions (May and November) might have introduced other confounding variables, such as differences in the academic calendar or varying levels of student motivation. We tried to mitigate this by ensuring a consistent experimental setup and providing identical instructions and incentives across both groups. However, any unobserved factors, such as personal interest in gaming or familiarity with testing tools, may still have affected the outcomes.

The \textit{external validity} of this study may be limited by the specific demographics of the participants, who were exclusively students at the \passau. The results may not generalize to students from other universities or to professional developers. Furthermore, the tasks and context within the game are highly tailored to the educational scenario, which may not reflect real-world testing and debugging scenarios. While the design of \toolname aims to provide realistic programming challenges, the controlled and gamified environment may differ significantly from actual software development settings. Thus, extrapolating the findings to broader populations or different contexts should be done cautiously.

\textit{Construct validity} could be impacted by how certain concepts and measurements were operationalized in the study. For example, metrics such as ``engagement'' and ``effectiveness in testing'' were derived from time spent on activities, test coverage, and survey responses. While these are common proxies, they may not capture the full depth of student engagement or the quality of their learning outcomes. Similarly, the absence of direct measurements for certain constructs, such as motivation or cognitive effort, might have limited our ability to fully understand the effects of \toolname on participants. We also relied on self-reported data for some aspects, such as prior experience, which introduces the risk of bias or inaccuracies.

	\section{Results}
	\subsection{RQ 1: How do students engage with \toolname?}

\begin{figure*}
  \centering
  	\begin{subfigure}[t]{0.33\textwidth}
		\centering
		\includegraphics[width=\textwidth]{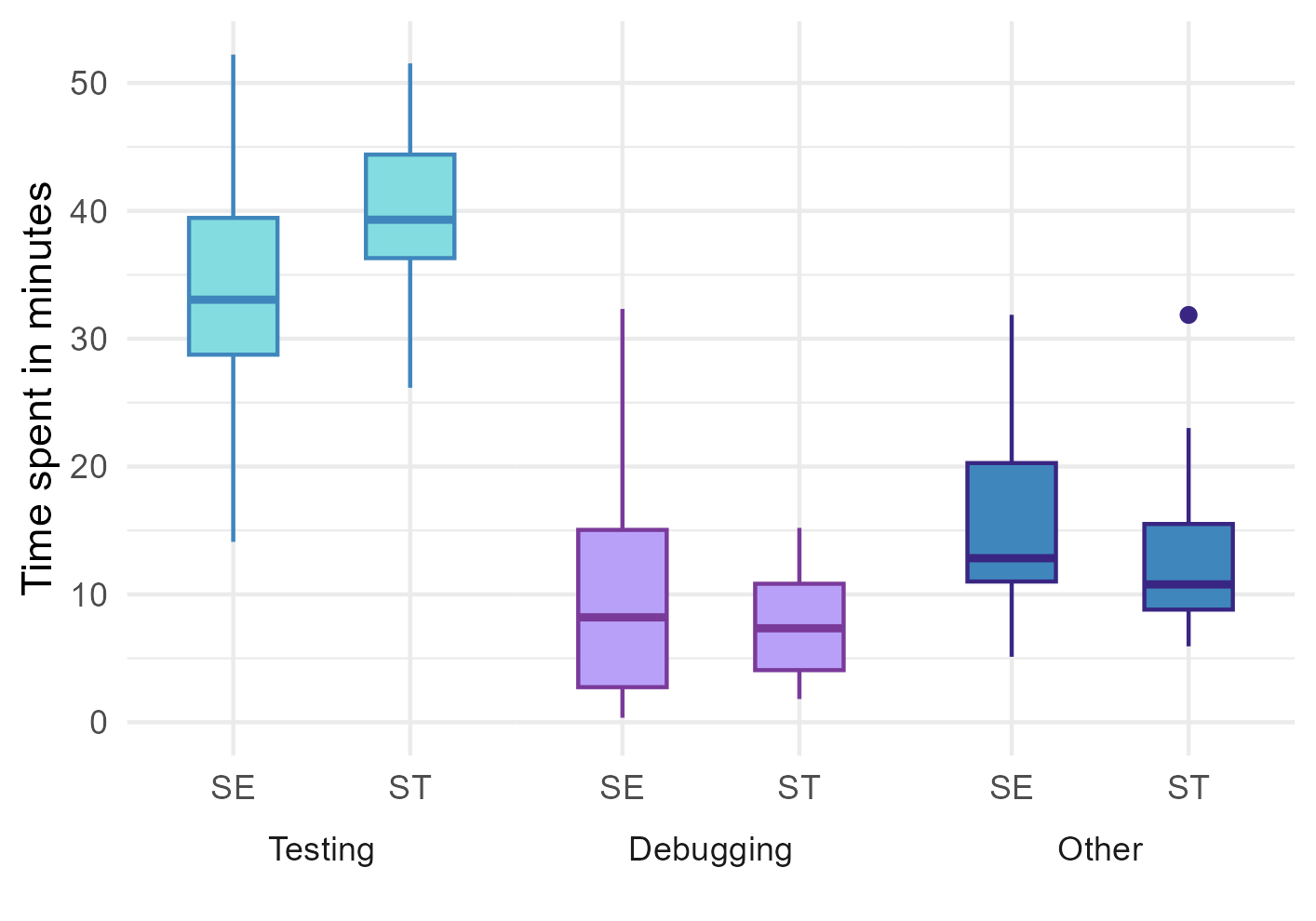}
		\vspace{-1em}
		\caption{Time spent per task for both groups}
		\label{fig:timetask}
              \end{subfigure}
	\hfill              
	\begin{subfigure}[t]{0.33\textwidth}
		\centering
		\includegraphics[width=\textwidth]{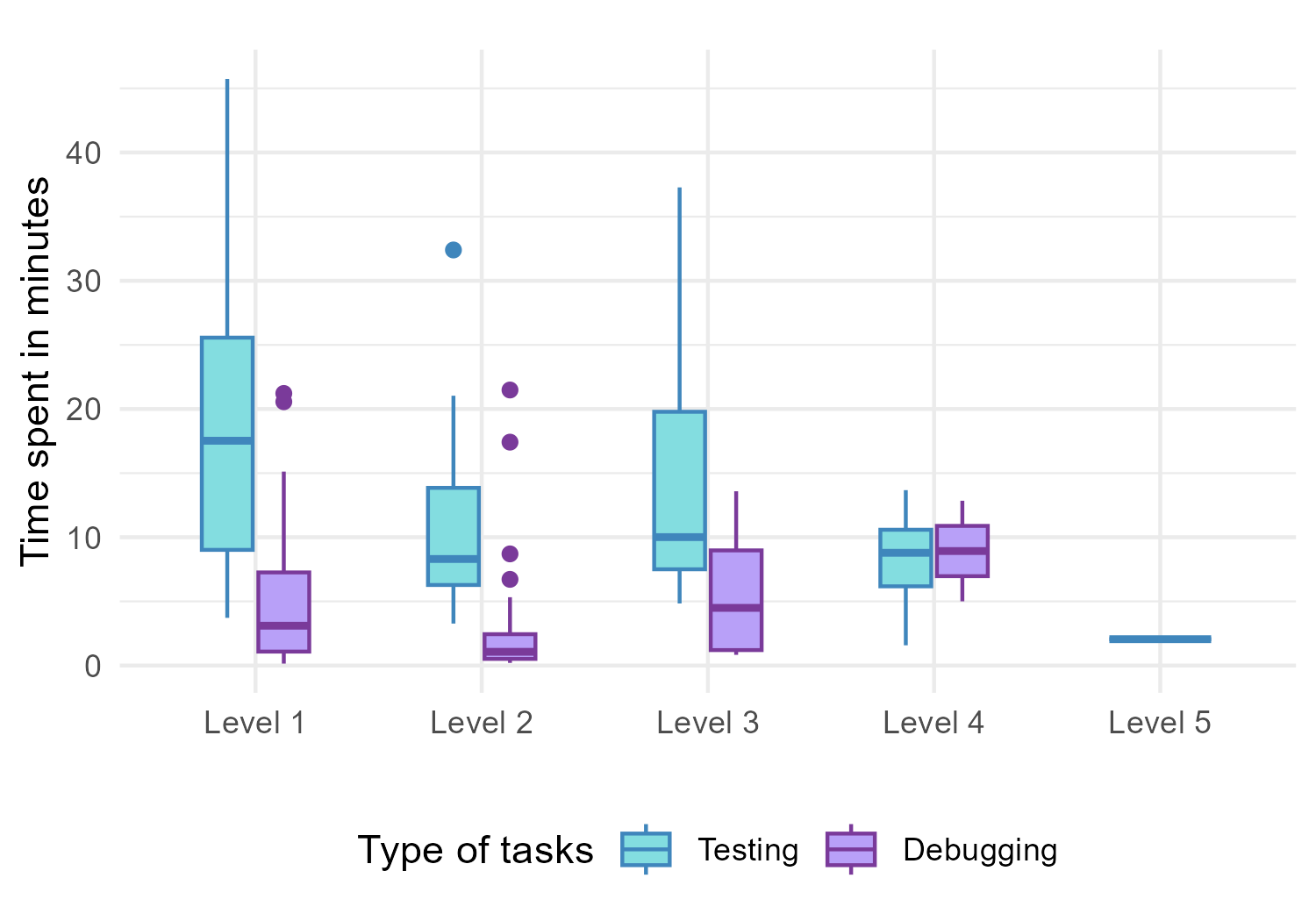}
		\vspace{-1em}
		\caption{Time spent per level for the SE group}
		\label{fig:timecomponentSE}
	\end{subfigure}
	\hfill
	\begin{subfigure}[t]{0.33\textwidth}
		\centering
		\includegraphics[width=\textwidth]{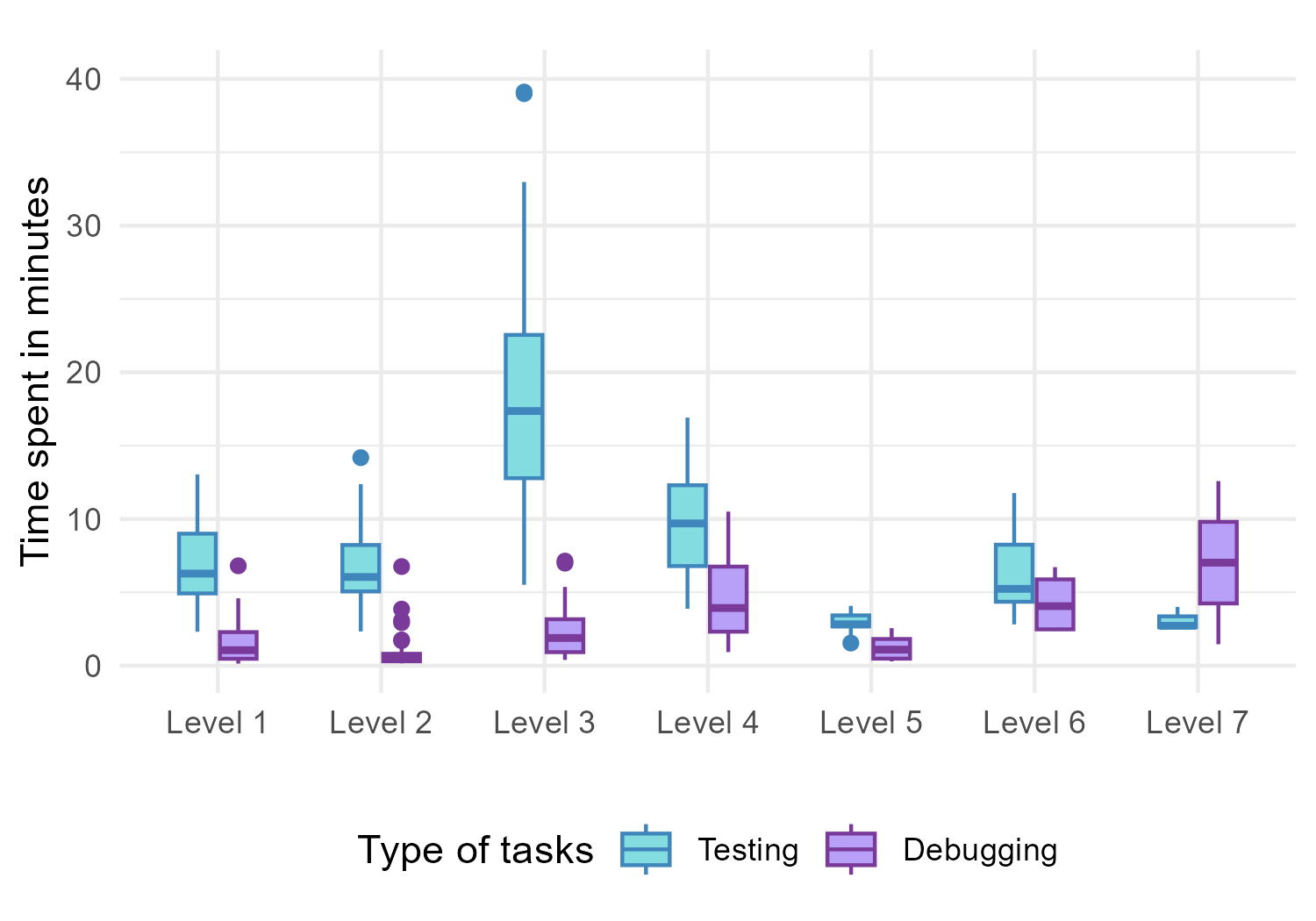}
		\vspace{-1em}
		\caption{Time spent per level for the ST group}
		\label{fig:timecomponentST}
	\end{subfigure}

	\caption{Time spent per level and task for both groups}
	\label{fig:timecomb}
\end{figure*}

\begin{figure}[t]
	\centering
	\includegraphics[width=0.8\linewidth]{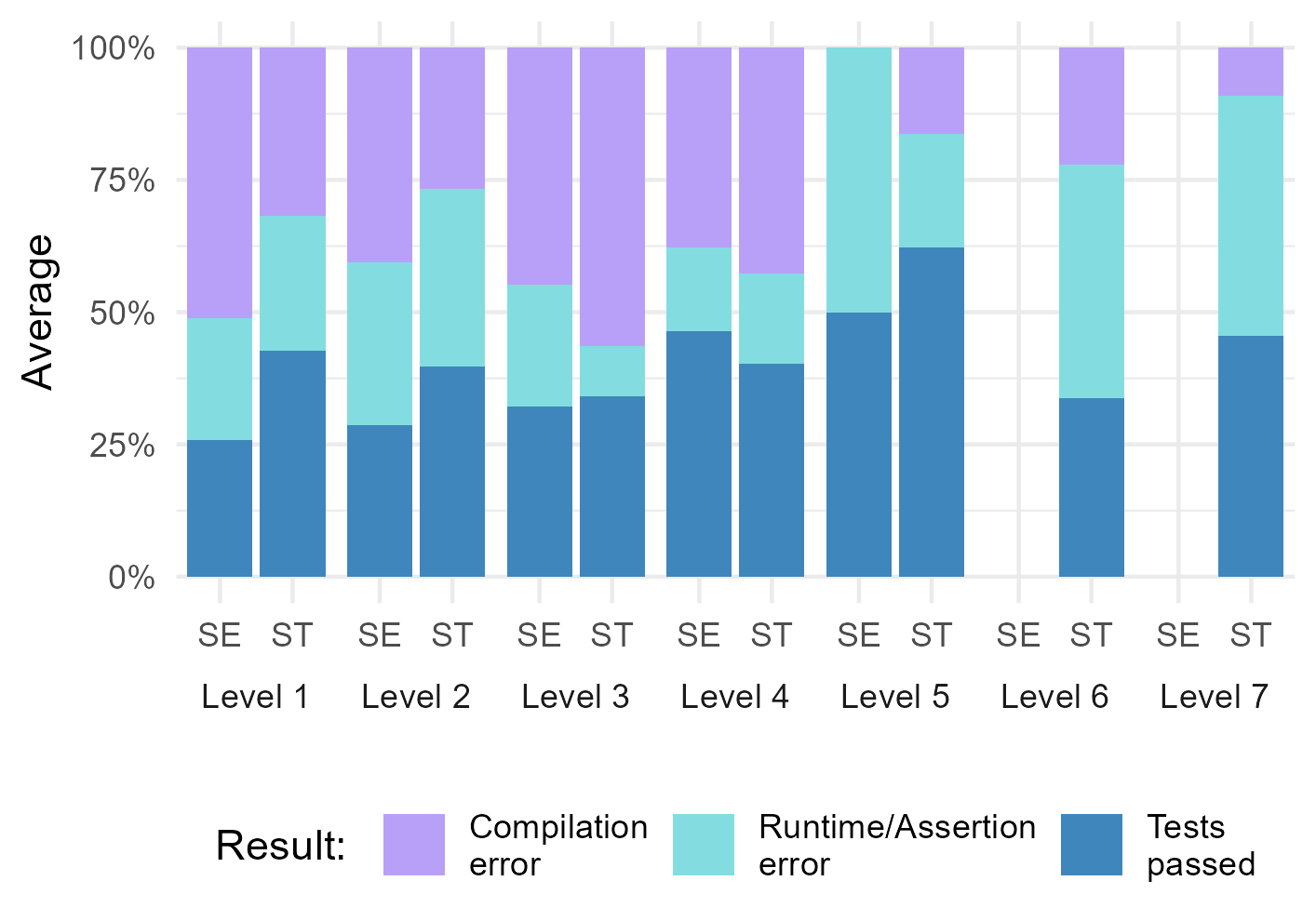}
	\caption{Attempts until activation for both groups}
	\label{fig:attempts}
\end{figure}

\Cref{fig:timecomb} shows how participants spent their time while playing. Overall, participants spent the most time on testing activities, with both groups averaging more than 35 minutes (\cref{fig:timetask}). In contrast, the time spent on debugging was significantly lower, averaging only about 8 minutes. This was also less than the time participants spent outside the editor engaging in activities such as exploring the map, interacting with objects in the rooms, and talking to the robot. Thus, participants primarily focused on writing tests and exploring the game, while less time was needed to locate and fix bugs.

The time spent on testing was consistently higher than the time spent debugging across most levels, as depicted in \cref{fig:timecomponentSE} and \cref{fig:timecomponentST}. An exception was level 7 for the ST group, where debugging required substantially more time than writing tests. This suggests that the bug in level 7 was particularly challenging to identify, likely due to the task's recursive nature, which demanded a deeper analysis of the execution flow. Participants also spent considerable time writing tests in level 3, indicating that this level was especially difficult for test creation, likely because it contained the highest number of lines and methods to cover.

When executing tests, participants encountered three possible outcomes: all tests passed, a compilation error occurred, or an assertion failed during execution; \Cref{fig:attempts} summarizes the distribution among these outcomes across all levels and groups. The success rate of test execution generally ranged between 25\% and 50\%, with the first three levels exhibiting the lowest success rates. The highest success rate was observed in level 5, suggesting it was the easiest level to understand. As participants progressed through the levels, the number of compilation errors decreased significantly after level 5, with only a few errors recorded. However, assertion errors increased after level 5, indicating that while participants became more comfortable writing code, they still encountered challenges in crafting correct assertions for their tests.

The ST group spent more time writing tests, averaging 41 minutes, compared to the SE group, which averaged 35 minutes (\cref{fig:timetask}, $p = 0.005$). Regarding level progression, the SE group primarily worked on levels 1 through 5 but did not complete level 5 (\cref{fig:timecomponentSE}). In contrast, the ST group worked on all seven levels, with most participants completing at least level 4 (\cref{fig:timecomponentST}), while most SE group participants completed only level 2. This disparity can be attributed to the greater programming experience of the ST group, as discussed in \cref{sec:participants}. In debugging activities, the ST group consistently outperformed the SE group, with significant differences observed in levels 1 ($p = 0.001$) and 2 ($p = 0.018$). However, direct comparisons between the groups are difficult because only the top-performing SE participants reached level 3, whereas all ST participants achieved this level. This suggests that in a 60-minute timeframe, only students with at least three years of programming experience are likely to complete all seven levels, while students with less experience require more time to do so.

\summary{RQ 1}{Participants focused primarily on testing and exploring the game, spending significantly less time on debugging activities. The ST group, benefiting from greater programming experience, progressed further through the levels and performed better in debugging, while less experienced SE participants required more time to advance.}

\subsection{RQ 2: How do students perform in testing activities with \toolname?}

\begin{figure*}
	\centering
	\begin{subfigure}[t]{0.33\textwidth}
		\centering
		\includegraphics[width=\textwidth]{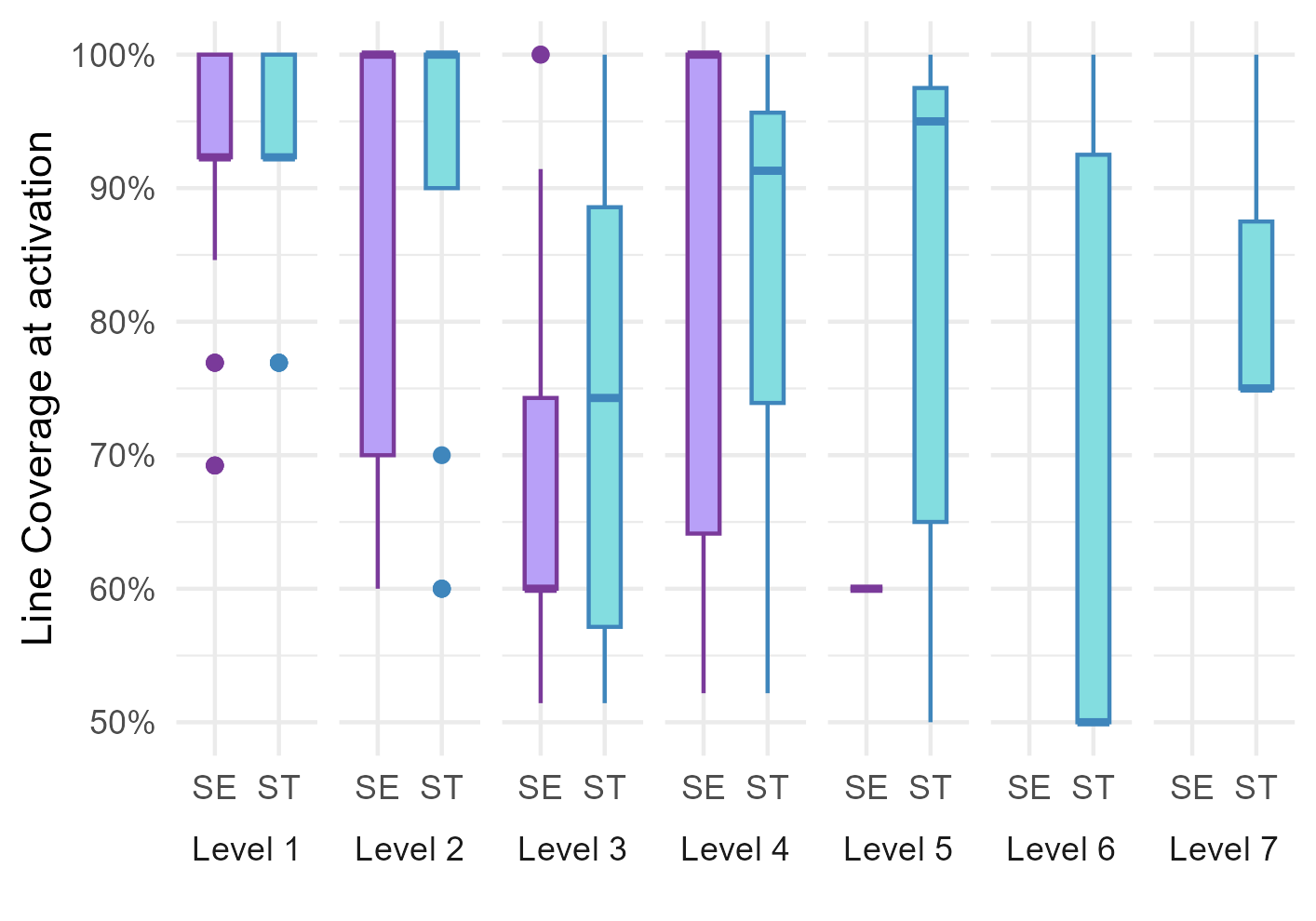}
		\vspace{-1em}
		\caption{Line coverage per level for both groups}
		\label{fig:linecoverage}
	\end{subfigure}
	\hfill
	\begin{subfigure}[t]{0.33\textwidth}
		\centering
		\includegraphics[width=\textwidth]{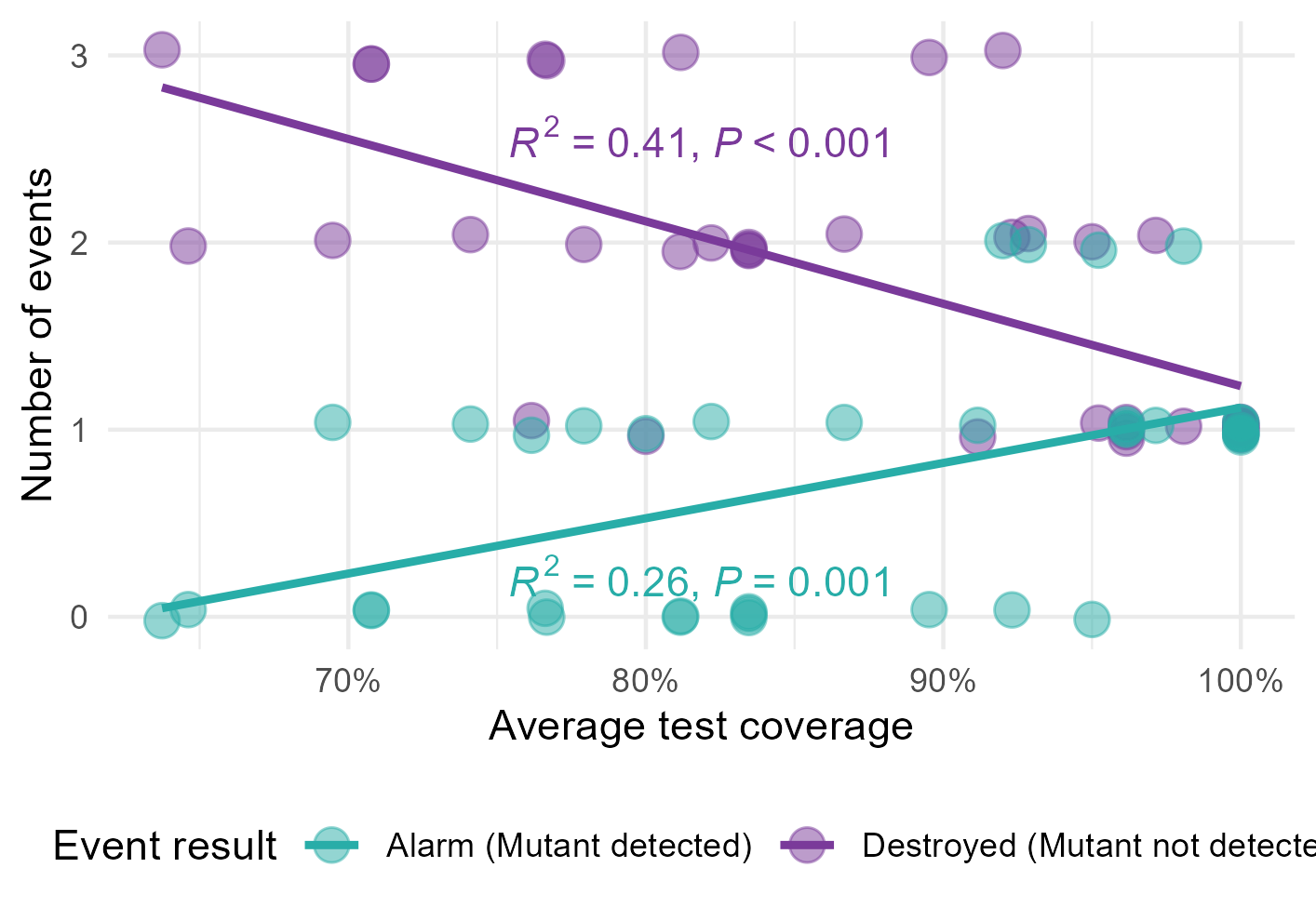}
		\vspace{-1em}
		\caption{Line coverage vs. target detected correlation for the SE group}
		\label{fig:coverageregressionSE}
	\end{subfigure}
	\hfill
	\begin{subfigure}[t]{0.33\textwidth}
		\centering
		\includegraphics[width=\textwidth]{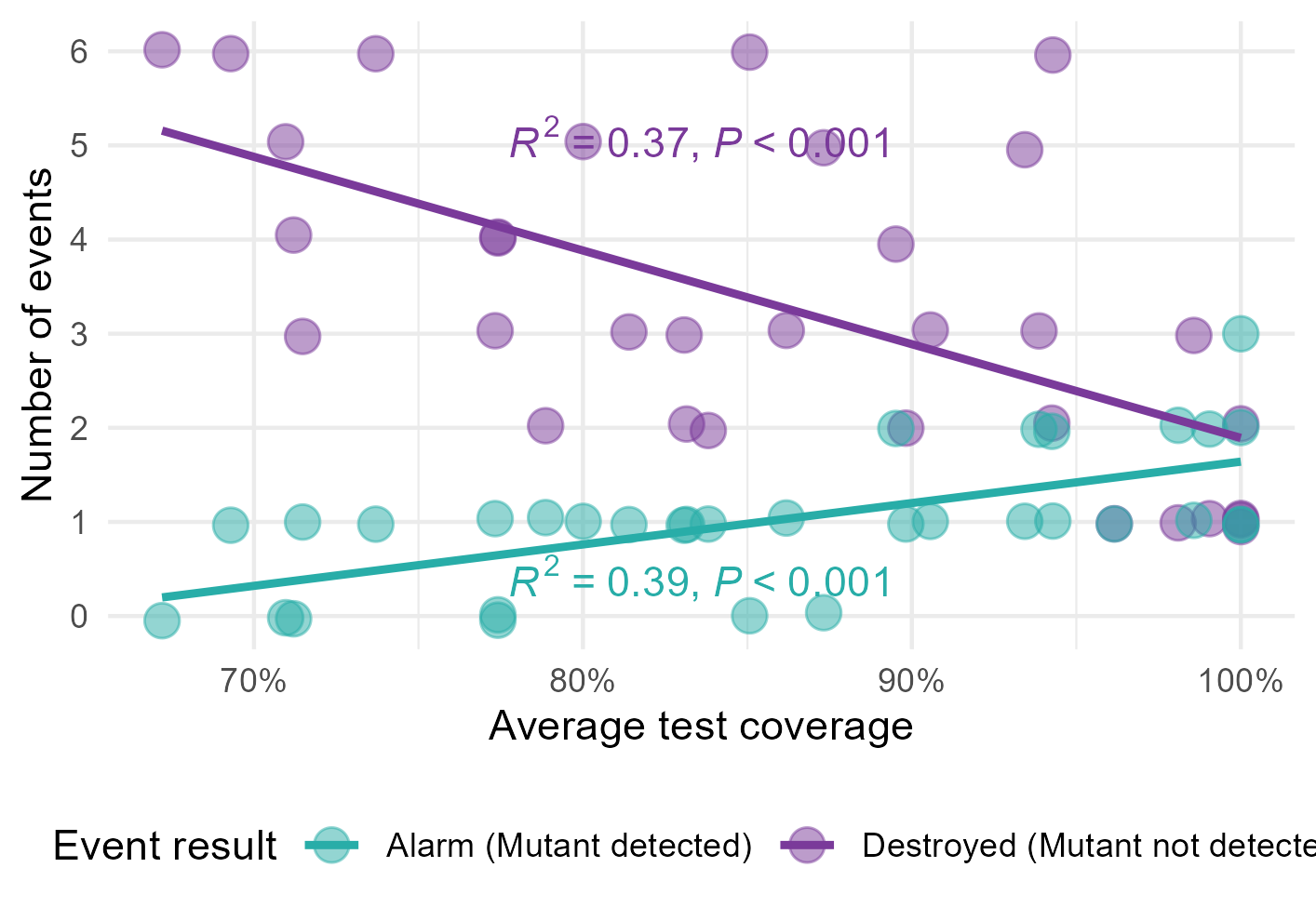}
		\vspace{-1em}
		\caption{Line coverage vs. target detected correlation for the ST group}
		\label{fig:coverageregressionST}
	\end{subfigure}
	
	\caption{Line coverage for both groups}
	\label{fig:coverage}
\end{figure*}

\begin{figure*}
	\centering
	\begin{subfigure}[t]{0.33\textwidth}
		\centering
		\includegraphics[width=\textwidth]{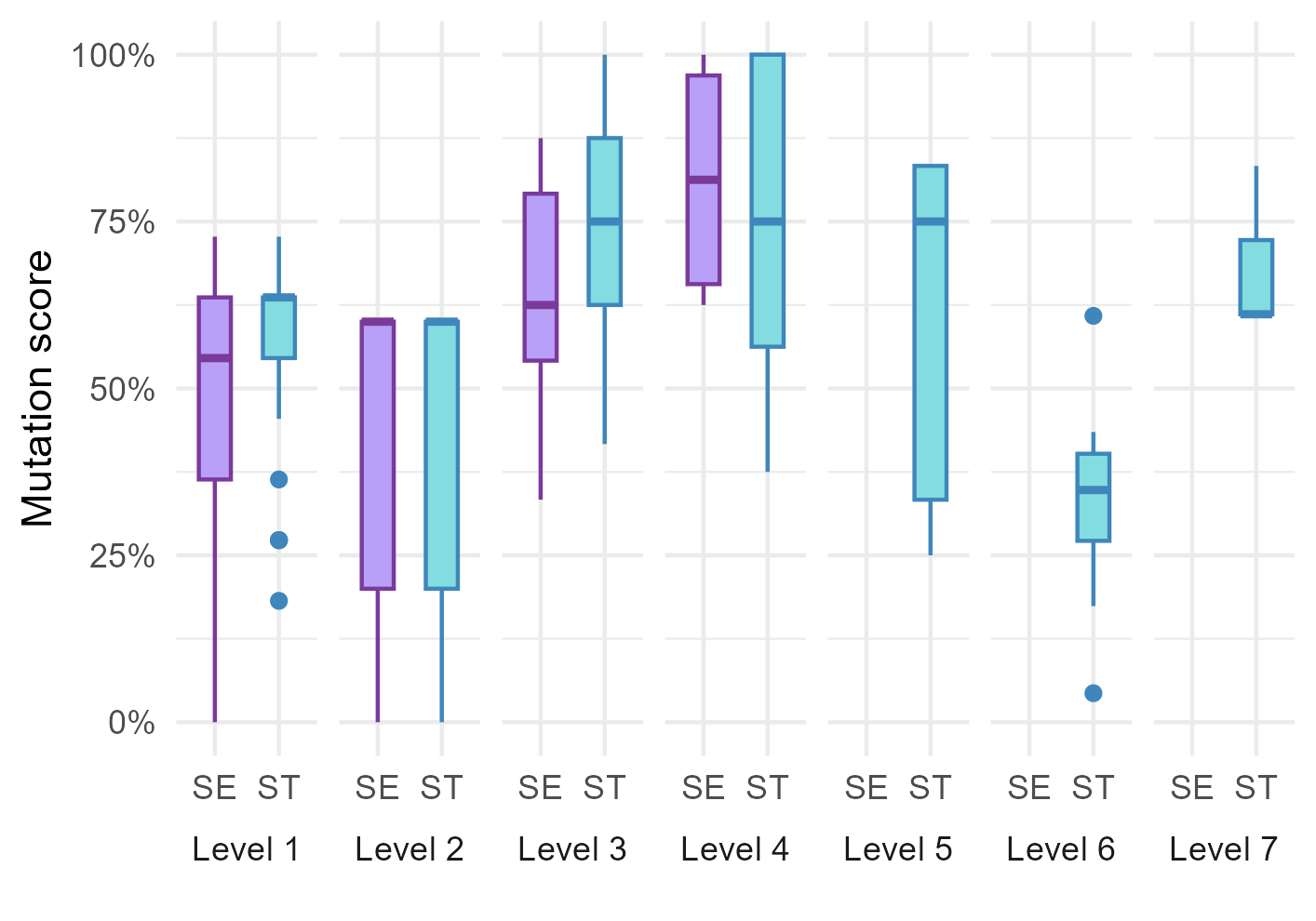}
		\vspace{-1em}
		\caption{Mutation score per level for both groups}
		\label{fig:mutationscore}
	\end{subfigure}
	\hfill
	\begin{subfigure}[t]{0.33\textwidth}
		\centering
		\includegraphics[width=\textwidth]{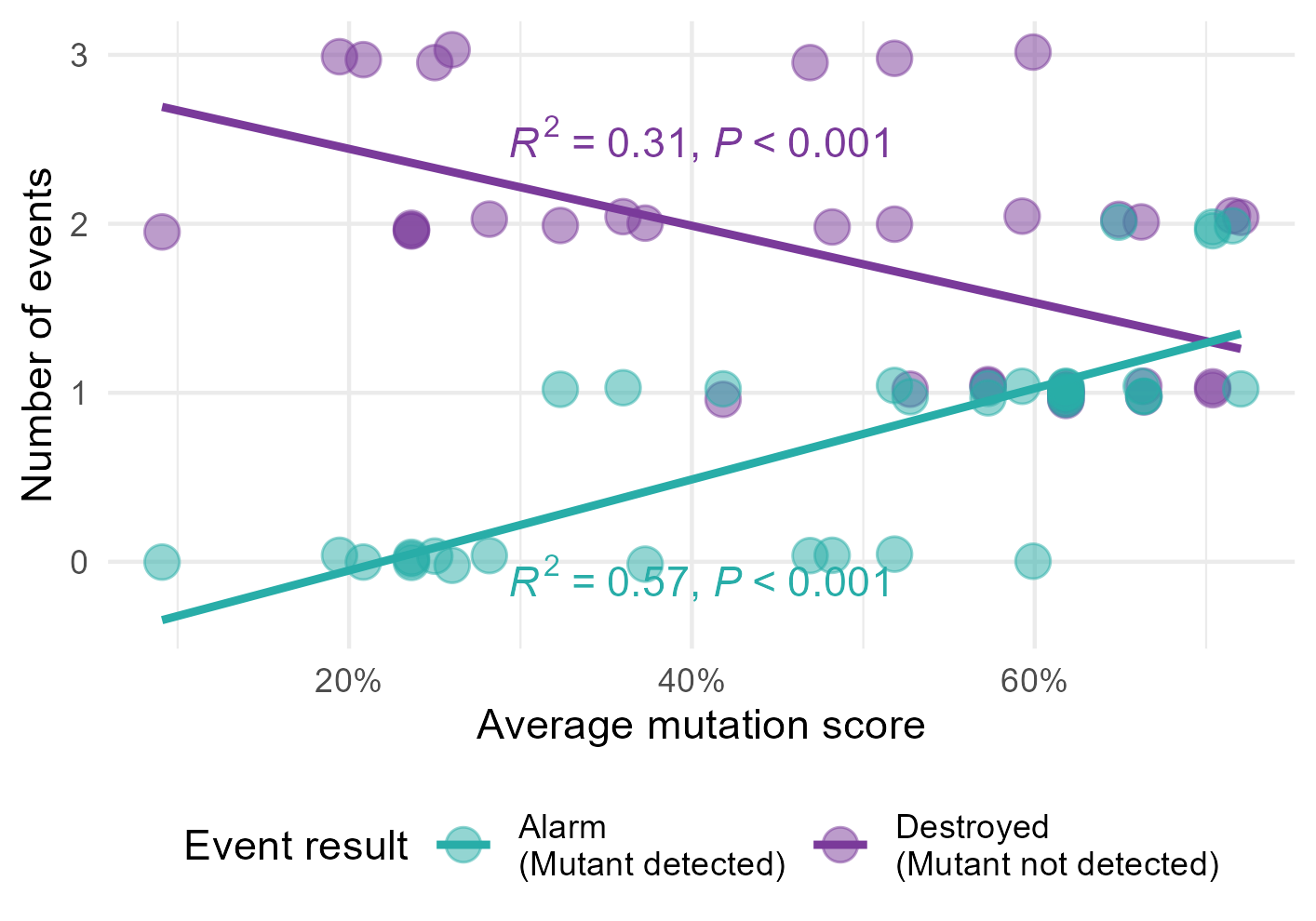}
		\vspace{-1em}
		\caption{Mutation score vs. target detected correlation for the SE group}
		\label{fig:mutationregressionSE}
	\end{subfigure}
	\hfill
	\begin{subfigure}[t]{0.33\textwidth}
		\centering
		\includegraphics[width=\textwidth]{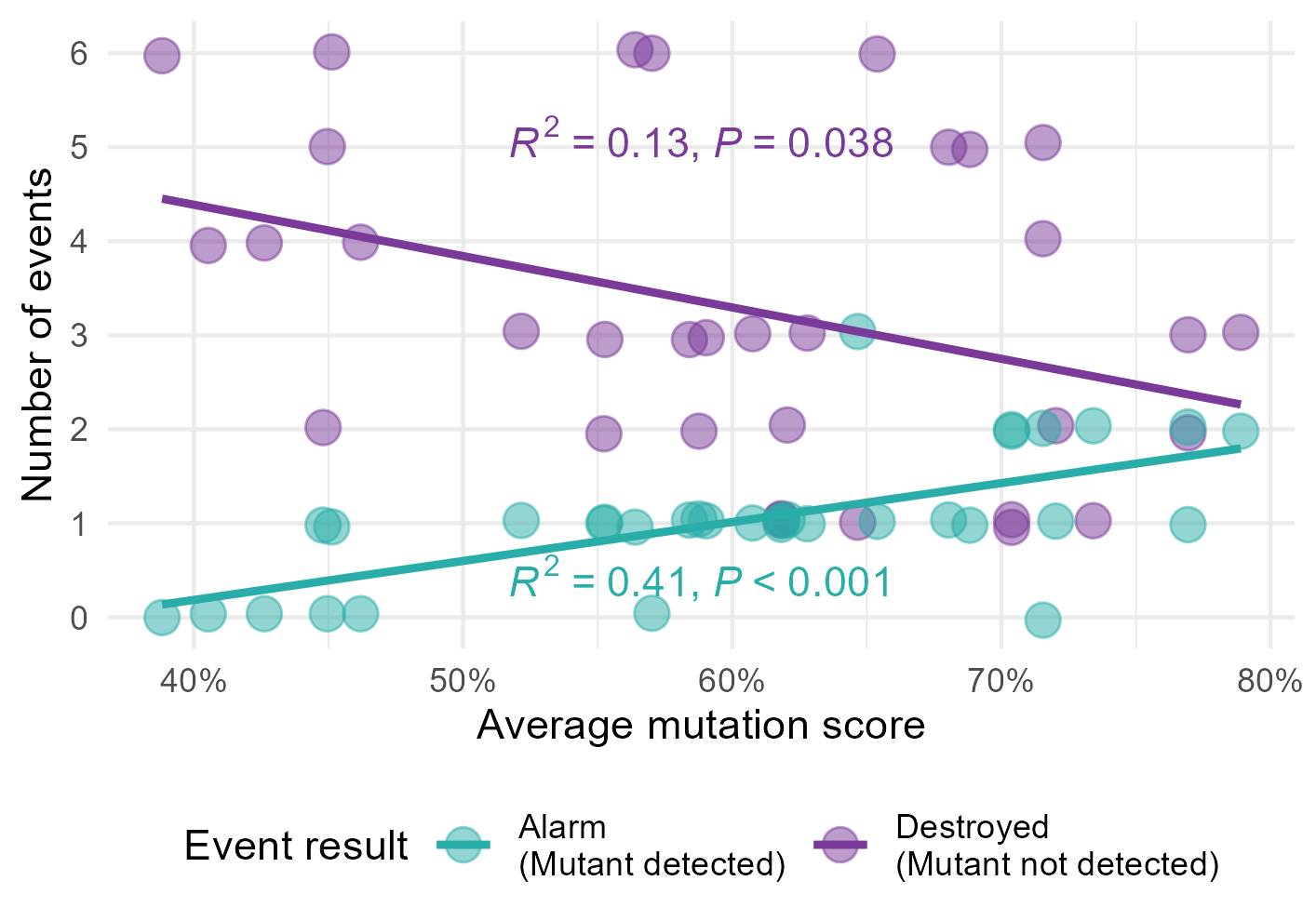}
		\vspace{-1em}
		\caption{Mutation score vs. target detected correlation for the ST group}
		\label{fig:mutationregressionST}
	\end{subfigure}
	
	\caption{Mutation score for both groups}
	\label{fig:mutation}
\end{figure*}

\begin{figure*}
	\centering
	\begin{subfigure}[t]{0.49\textwidth}
		\centering
		\includegraphics[width=\textwidth]{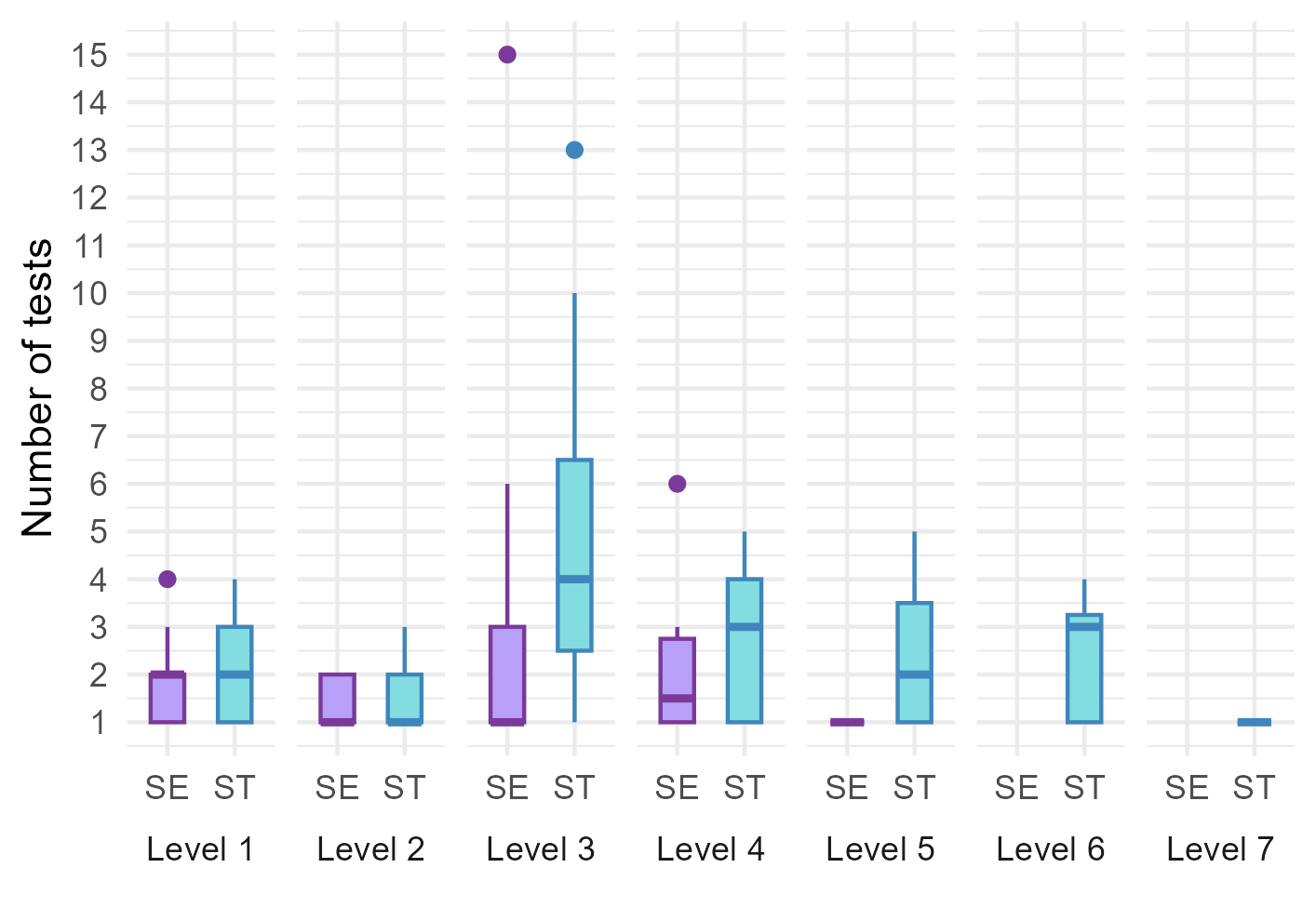}
		\vspace{-1em}
		\caption{Number of tests per level for both groups}
		\label{fig:tests}
	\end{subfigure}
	\hfill
	\begin{subfigure}[t]{0.49\textwidth}
		\centering
		\includegraphics[width=\textwidth]{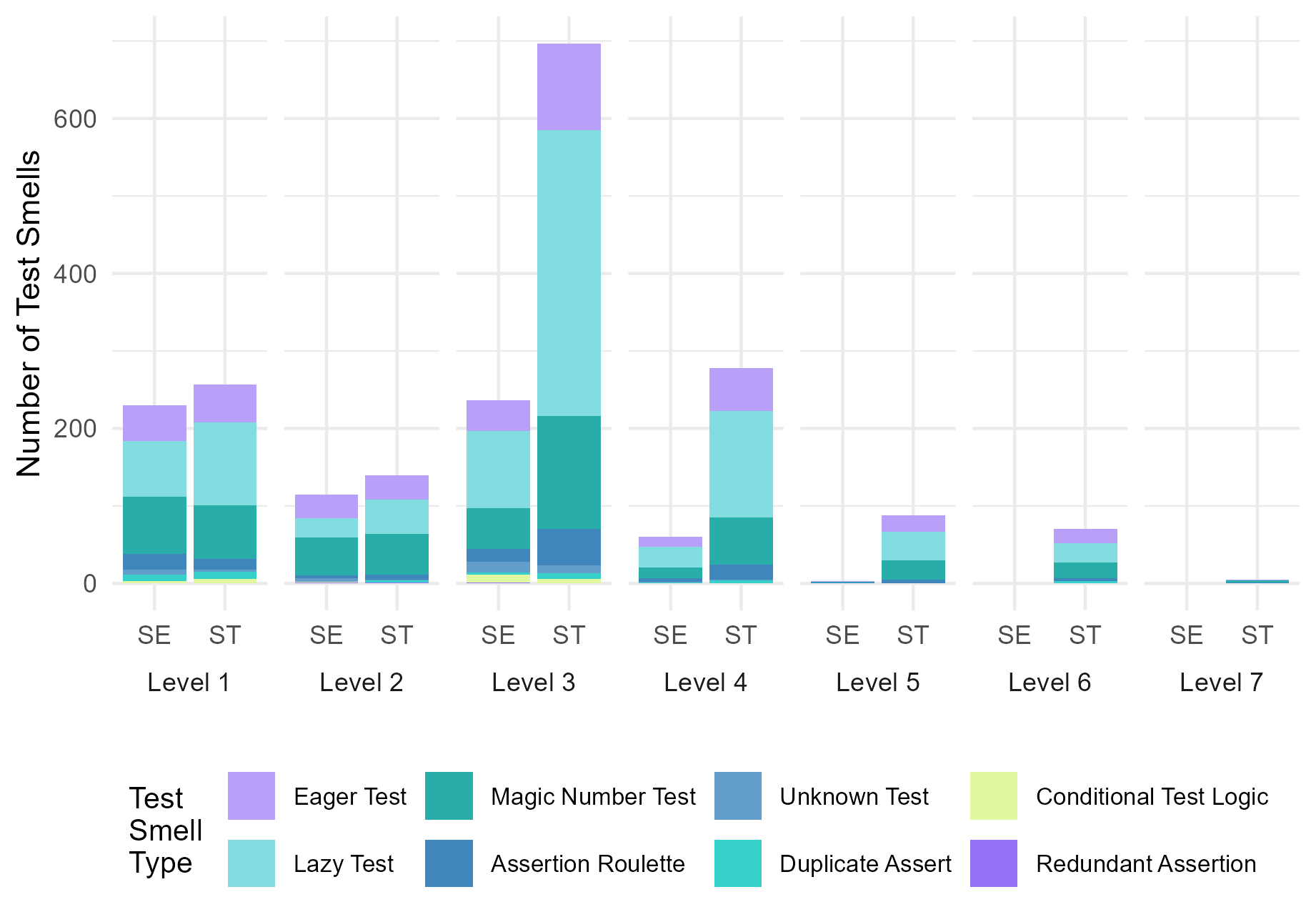}
		\vspace{-1em}
		\caption{Number of smells per level for both groups}
		\label{fig:smells}
	\end{subfigure}
	
	\caption{Number of tests and smells per level for both groups}
	\label{fig:testssmells}
\end{figure*}

\Cref{fig:coverage} summarizes line coverage achieved by the players. Considering the line coverage per level (\cref{fig:linecoverage}), both groups performed similarly across levels, consistently surpassing the 50\% activation threshold. In level 1, participants achieved an average of 83\% coverage, with nearly 100\% in level 2. While coverage varied more in the later levels due to increased task complexity, it always exceeded the threshold, indicating participants wrote more tests than required and were committed to test creation.
Although expectedly lower than coverage levels, the mutation scores across levels (\cref{fig:mutationscore}) consistently exceeded 50\% except in level 6, which scored about 35\%, highlighting its difficulty in eliminating mutants. Mutation scores peaked in levels 3, 4, and 5, despite participants not explicitly focusing on mutation elimination. Although the mutation scores were generally good, there is clearly room for improvement, demonstrating that the levels are not trivial to test.

The highest number of tests was written for level 3 (\cref{fig:tests}), aligning with its complexity as the level containing the most lines and methods. However, the abundance of tests in level 3 also introduced the highest number of test smells (\cref{fig:smells}). The most common test smell type is \textit{Lazy Test}, where participants wrote multiple tests calling the same method within a component. This is unsurprising, as it reflects a fine-grained testing approach. Other frequently observed test smell types include \textit{Magic Number Test}, where players directly used hardcoded numbers in their assertions instead of defining them as constants, and \textit{Eager Test}, where a single test calls multiple methods in a component, indicating a lack of isolation. These issues highlight opportunities for improvement, which could be addressed in future versions.

In terms of detecting the target bugs, the tests written for the first level were the most successful, identifying the bug in approximately 75\% of cases for the ST group and 63\% for the SE group. Beyond the first level, the trend shifted, with tests generally failing to detect the bugs. This suggests that focusing primarily on achieving 50\% line coverage was insufficient for bug detection, which can also be seen in terms of mutation score. However, analysis of the data shows that higher coverage increases the likelihood of detecting bugs. For both groups, there is a weak positive correlation between coverage and detection of seeded bugs (\cref{fig:coverageregressionSE} and \cref{fig:coverageregressionST}). Conversely, the correlation for failing to detect seeded bugs decreases, with a moderate negative correlation observed in the SE group and a weak negative correlation in the ST group. A similar trend is observed with mutation scores: The higher the mutation score, the greater the probability of detecting the seeded bug, with a moderate positive correlation for both groups (\cref{fig:mutationregressionSE} and \cref{fig:mutationregressionST}). At the same time, the likelihood of not detecting the bug decreases, showing a weak negative correlation in the SE group and a very weak negative correlation in the ST group. This demonstrates that success in the game is clearly related to being able to apply adequate testing.

Both groups exceeded the 50\% coverage threshold consistently across levels, with the ST group generally achieving higher coverage due to their additional testing time and experience, although the differences were not statistically significant except in levels 1 ($p = 0.026$) and 3 ($p = 0.048$). At level 4, the SE group achieved a slightly higher mutation score, though this result should be interpreted cautiously, as only the best SE players reached this level. Overall, the ST group wrote more tests than the SE group on average, but this did not always translate to higher efficiency, as the SE group often achieved similar outcomes with fewer tests. However, the higher number of tests by the ST group contributed to better coverage and mutation scores, particularly at level 3, where they also recorded significantly more test smells (nearly 700 vs. fewer than 250, $p = 0.015$). This suggests that while the ST group was more thorough, they also faced challenges in maintaining test quality.

\summary{RQ 2}{Both groups consistently exceeded the 50\% line coverage threshold, with the more experienced ST group generally achieving higher coverage and mutation scores. However, while the ST group wrote more tests and achieved better metrics, they also introduced more test smells, indicating challenges in maintaining test quality, especially at complex levels.}

\subsection{RQ 3: How do students perform in debugging activities with \toolname?}

\begin{figure*}
	\centering
	\begin{subfigure}[t]{0.4\textwidth}
		\centering
		\includegraphics[width=\textwidth]{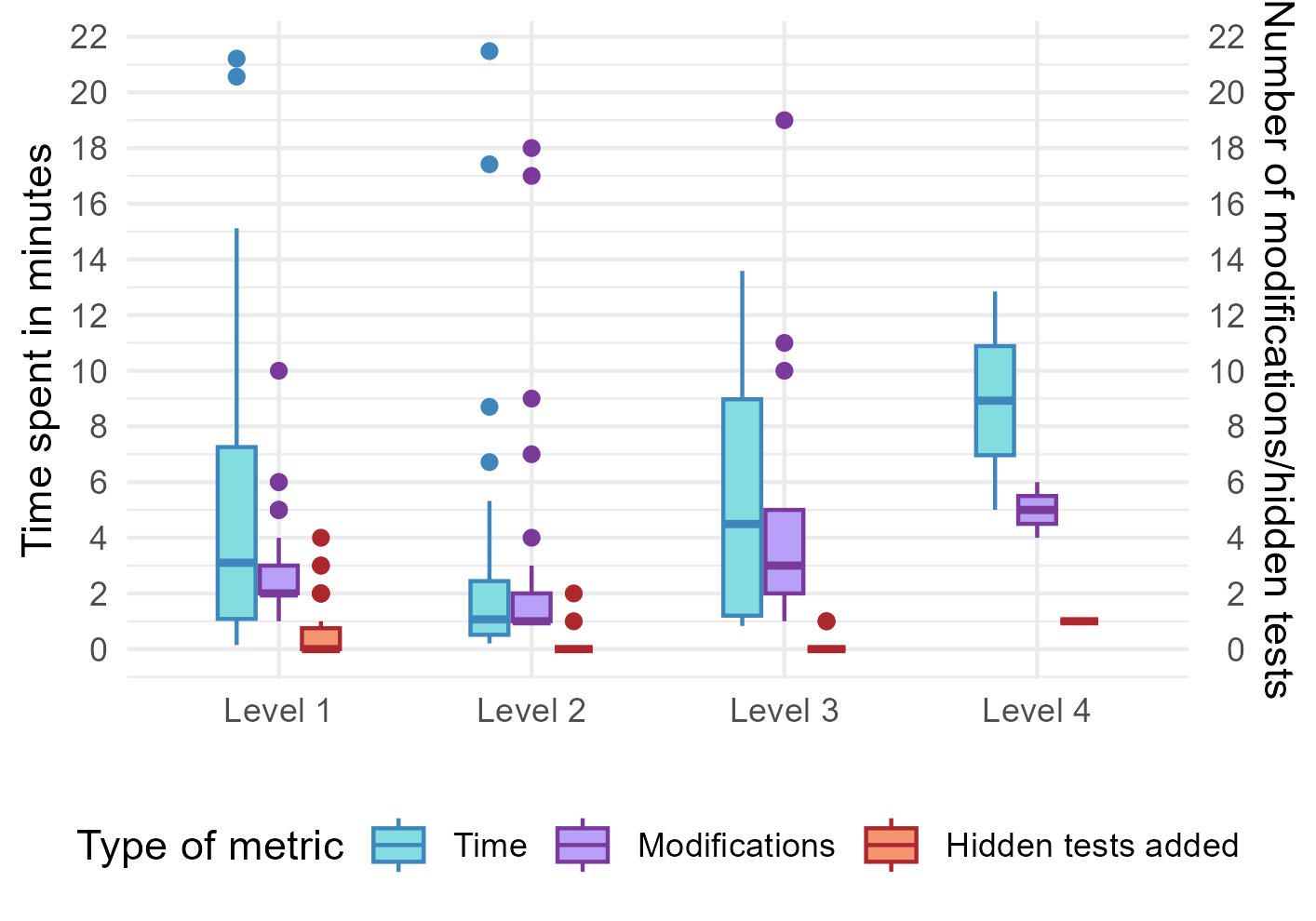}
		\vspace{-1em}
		\caption{Debugging performance per level for the SE group}
		\label{fig:debugSE}
	\end{subfigure}
	\hfill
	\begin{subfigure}[t]{0.4\textwidth}
		\centering
		\includegraphics[width=\textwidth]{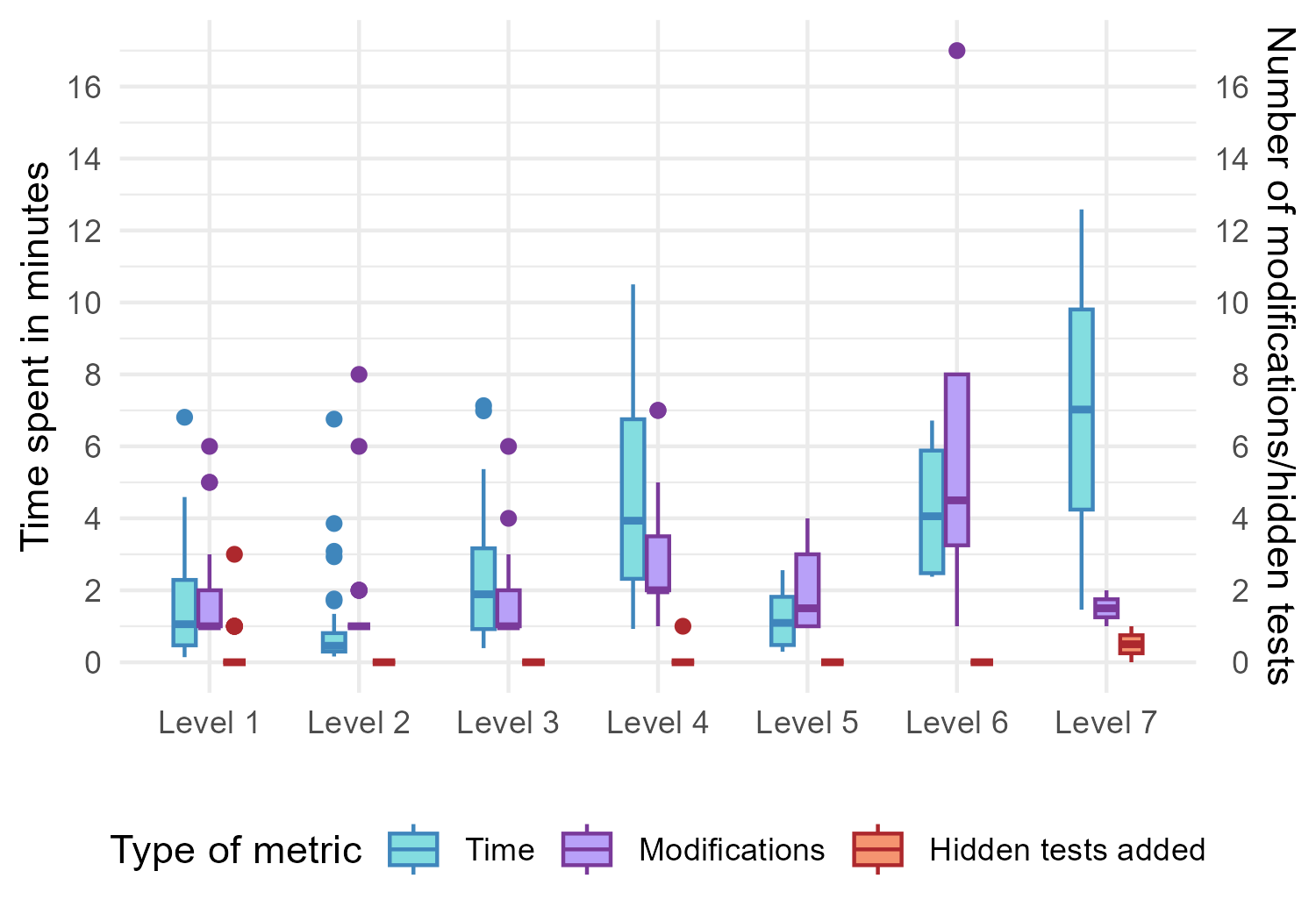}
		\vspace{-1em}
		\caption{Debugging performance per level for the ST group}
		\label{fig:debugST}
	\end{subfigure}
	
	\caption{Debugging performance per level for both groups}
	\label{fig:debugcomb}
\end{figure*}


\Cref{fig:debugcomb} summarizes for each level how much time the students spent for debugging, how many attempts they required in order to fix the bug, and how often they accidentally introduced new bugs causing hidden tests to be added to their test suite. The first level likely required some additional time to familiarize with this aspect of the game, while the second level appears to be the easiest overall. From then onward, the difficulty seems to gradually increase, although interestingly level five, which was not completed by any participants of the SE group (\cref{fig:debugSE}), seems easy for the ST group (\cref{fig:debugST}). This is also reflected by the number of attempts required across the levels, except for level 7 where students required  substantial time understanding the recursive component but made minimal modifications, introducing new bugs in the process, which highlights the level's difficulty.
Besides this, new bugs were introduced infrequently, with some bugs appearing in levels 1 and 4. Neither group frequently used the print statement debugging feature, with the ST group using it six times and the SE group 17 times. This low usage suggests participants either overlooked the feature or found it inconvenient, despite prior instructions. 

Significant differences between groups were observed in time spent on levels 1 ($p < 0.001$) and 2 ($p = 0.009$), and in code modification activities for levels 1 ($p = 0.007$), 2 ($p = 0.015$) and 3 ($p = 0.002$), with the ST group demonstrating more caution during modifications, likely due to greater programming experience. Differences in bug introduction were significant only in levels 3 ($p = 0.043$) and 4 ($p = 0.002$), where the SE group was more prone to do so.

\summary{RQ 3}{Participants initially focused more on understanding the code and executing tests but shifted to code modifications in later levels, due to uncertainty about fixing bugs. The ST group was more cautious and introduced fewer bugs than the SE group, with significant differences observed in specific levels, while the low usage of the print debugging feature suggests better debugging features (e.g., breakpoints, stepping) may be needed.}

\subsection{RQ 4: How do students perceive \toolname?}

\begin{figure}[t]
	\centering
	\includegraphics[width=0.9\linewidth]{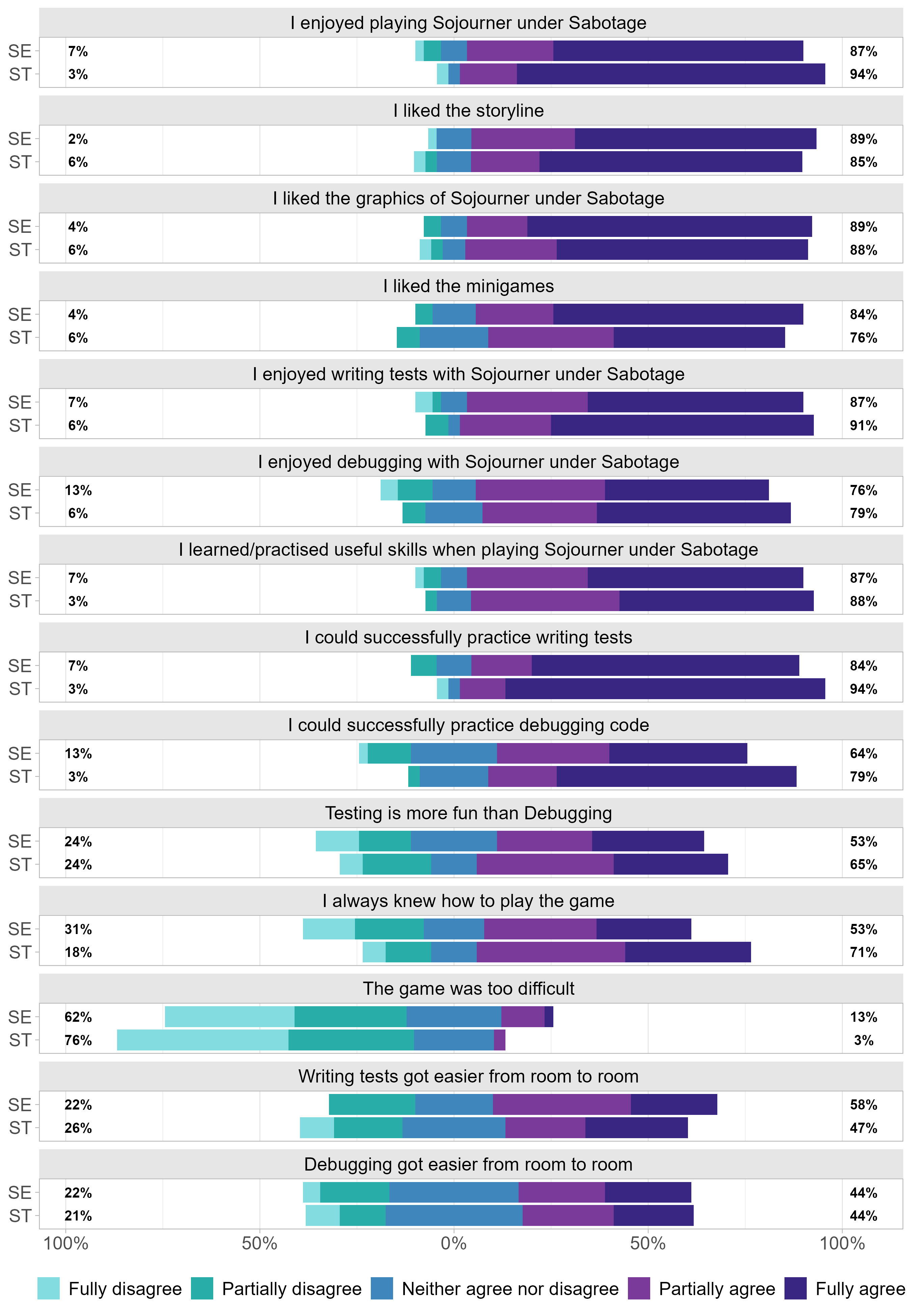}
	\caption{Participants' feedback from both groups (SE, ST)}
        \vspace{-1em}
	\label{fig:survey}
\end{figure}

The participants' feedback after playing \toolname was overwhelmingly positive in both groups, as shown in \cref{fig:survey}. Over 80\% of participants enjoyed the game, appreciating the storyline, graphics, and mini-games. Approximately 90\% of participants in both groups enjoyed writing tests, and over 75\% also found debugging enjoyable. Most participants agreed that they learned useful skills and successfully practiced testing while playing.  

However, there was a slight difference in responses regarding practicing debugging skills: 79\% of the ST group agreed, compared to only 64\% of the SE group. This suggests that less experienced students may have found debugging more challenging than testing. Interestingly, only about half of the SE group stated that testing was more fun than debugging, compared to 65\% of the ST group, indicating mixed feelings among the less advanced participants.  

The hypothesis that the SE group occasionally felt overwhelmed is supported by the fact that only about half of them consistently knew how to play the game, compared to over 70\% of the ST group. Additionally, more than 75\% of the ST group felt the game was not too difficult, whereas only 64\% of the SE group agreed. Interestingly, less than half of the ST group felt that testing became easier as the levels progressed, while almost 60\% of the SE group noticed an improvement. This suggests that the ST group perceived the difficulty across levels as more consistent, whereas the SE group found testing easier over time. For debugging, 44\% of both groups felt it became easier as the game progressed, indicating a shared perception of the difficulty of debugging tasks.

\summary{RQ 4}{The feedback from both groups was overwhelmingly positive, suggesting the game is equally enjoyable for complete beginners and more advanced learners.
%
  While the SE group found testing easier over time, they struggled more with debugging, whereas the ST group perceived the difficulty as consistent across levels and demonstrated greater confidence throughout.}

	\section{Related Work}
	\subsection{Serious Games in Testing}

The Testing Game~\cite{DBLP:conf/fie/ValleTBM17} consists of three levels, each covering a specific testing technique: functional, structural, and defect-based tests. Players solve puzzles and quizzes related to these techniques without writing tests, using a bubble sort implementation in Java as the CUT. The game is single-player and browser-based. Its evaluation involved 15 participants with prior knowledge of software testing, focusing solely on the students' perspectives.
In contrast, \toolname requires players to write tests, offering hands-on practice with real-world testing frameworks. While this restricts the number of testing techniques addressed, it provides practical experience. Unlike Testing Game, its evaluation targeted users with minimal or no prior testing knowledge, creating a genuine learning scenario and assessing their suitability for its audience.

Unlike Testing Game, Code Defenders~\cite{DBLP:conf/icse/CleggRF17,DBLP:conf/sigcse/FraserGKR19} is a multiplayer browser-based game where players compete by writing unit tests in Java using JUnit and creating mutations of the CUT, with a leaderboard showcasing top players. It is used in lectures, and its evaluation includes student feedback, in-game performance data, and the correlation between game performance and grades.
Key criticisms of Code Defenders include limitations of the code editor, performance issues, and repetitive gameplay. These concerns were considered in \toolname, which shares the mechanic of writing JUnit tests. Efforts were made to improve the code editor’s usability and ensure the CUTs were varied to reduce monotony.

Code Critters~\cite{DBLP:conf/icst/StraubingerCF23,DBLP:conf/icst/StraubingerBF24} is a Tower Defense-inspired serious game teaching software testing. Players place magic portals to protect healthy creatures from mutants, mimicking testing tasks like selecting test data and creating oracles using a block-based programming language. Designed for younger learners, the game engagingly introduces testing concepts. A study with 40 children showed high engagement and enjoyment, with many demonstrating meaningful interaction and some choosing to play voluntarily in their free time.
In contrast, \toolname is designed for undergraduate students rather than secondary education. While Code Critters uses a block-based language to simplify testing concepts for children, \toolname employs a serious game to engage students with prior knowledge, helping them learn and practice testing.

\subsection{Serious Games in Debugging}

Gidget~\cite{DBLP:conf/vl/Lee14} teaches programming through debugging, where players fix partially correct Python-like code to guide a robot in a factory-cleanup scenario. Designed for beginners, it successfully introduces programming concepts, using principles like learning by debugging, story-driven tasks, clear goals, and in-game help.  
While \toolname does not aim to teach programming, it draws on Gidget’s success, incorporating story-driven tasks, clear objectives, and the concept of computers as fallible components.

RoboBUG~\cite{DBLP:conf/icer/MiljanovicB17} is a game teaching debugging techniques like code tracing, print statements, and divide-and-conquer through provided C++ code. Players navigate the code environment to locate bugs, guided by a storyline about saving the world from alien bugs. Targeted at first-year CS students, evaluations showed improved debugging skills, especially for those with less prior knowledge, though frustration persisted.
Similarly, \toolname focuses on bug detection in provided code but currently lacks features like breakpoints or step-by-step execution.

	\section{Conclusions}
	In this paper, we introduced \toolname, a browser-based serious game designed to teach software testing and debugging through an engaging and interactive experience. Our evaluation involving 79 students demonstrates that the game effectively motivates students to practice these essential skills.
Overall, more than 80\% of participants enjoyed the game, praising its storyline, visuals, and educational value. These positive results were observed both with complete beginner as well as more advanced students, showing that both target groups can benefit from playing the game.

We also identified some challenges that we will aim to address in future iterations of \toolname. The ST group exhibited a higher number of test smells, reflecting areas for improvement in test-writing practices, and both groups rarely utilized debugging features like print statements, suggesting these tools were either overlooked or difficult to use. Our plans include improving debugging tools by incorporating features such as step-through execution and breakpoints to reduce the reliance on basic print debugging. Additionally, the game will integrate better guidance to encourage quality test-writing practices, addressing common issues such as \textit{Lazy Tests} and \textit{Magic Number Tests}. To expand its educational scope, adaptive difficulty mechanisms will be explored to provide tailored challenges, ensuring engagement for students with different skill levels. Finally, the game’s effectiveness will be validated through evaluations with a more diverse audience, including students from other institutions and professional developers, to assess its applicability across various contexts.

	To support replications and further
	research on \toolname, all our
	experiment material is available at:
	\begin{center}
		\url{https://doi.org/10.6084/m9.figshare.28838414}
	\end{center}
	\toolname can be played at:
	\begin{center}
		\url{https://sojourner-under-sabotage.se2.fim.uni-passau.de/}
	\end{center}
	
	\balance
	
	\bibliographystyle{ACM-Reference-Format}
	\bibliography{bib}


\begin{thebibliography}{34}


\ifx \showCODEN    \undefined \def \showCODEN     #1{\unskip}     \fi
\ifx \showDOI      \undefined \def \showDOI       #1{#1}\fi
\ifx \showISBNx    \undefined \def \showISBNx     #1{\unskip}     \fi
\ifx \showISBNxiii \undefined \def \showISBNxiii  #1{\unskip}     \fi
\ifx \showISSN     \undefined \def \showISSN      #1{\unskip}     \fi
\ifx \showLCCN     \undefined \def \showLCCN      #1{\unskip}     \fi
\ifx \shownote     \undefined \def \shownote      #1{#1}          \fi
\ifx \showarticletitle \undefined \def \showarticletitle #1{#1}   \fi
\ifx \showURL      \undefined \def \showURL       {\relax}        \fi
\providecommand\bibfield[2]{#2}
\providecommand\bibinfo[2]{#2}
\providecommand\natexlab[1]{#1}
\providecommand\showeprint[2][]{arXiv:#2}

\bibitem[\protect\citeauthoryear{Blanco, Trinidad, Cabal, Calder{\'{o}}n, Ruiz,
  and Tuya}{Blanco et~al\mbox{.}}{2023}]%
        {DBLP:journals/jss/BlancoTCCRT23}
\bibfield{author}{\bibinfo{person}{Raquel Blanco}, \bibinfo{person}{Manuel
  Trinidad}, \bibinfo{person}{Mar{\'{\i}}a Jos{\'{e}}~Su{\'{a}}rez Cabal},
  \bibinfo{person}{Alejandro Calder{\'{o}}n}, \bibinfo{person}{Mercedes Ruiz},
  {and} \bibinfo{person}{Javier Tuya}.} \bibinfo{year}{2023}\natexlab{}.
\newblock \showarticletitle{Can gamification help in software testing
  education? Findings from an empirical study}.
\newblock \bibinfo{journal}{\emph{J. Syst. Softw.}}  \bibinfo{volume}{200}
  (\bibinfo{year}{2023}), \bibinfo{pages}{111647}.
\newblock
\urldef\tempurl%
\url{https://doi.org/10.1016/J.JSS.2023.111647}
\showDOI{\tempurl}


\bibitem[\protect\citeauthoryear{Bonar and Soloway}{Bonar and Soloway}{1985}]%
        {DBLP:journals/hhci/BonarS85}
\bibfield{author}{\bibinfo{person}{Jeffrey Bonar} {and} \bibinfo{person}{Elliot
  Soloway}.} \bibinfo{year}{1985}\natexlab{}.
\newblock \showarticletitle{Preprogramming Knowledge: {A} Major Source of
  Misconceptions in Novice Programmers}.
\newblock \bibinfo{journal}{\emph{Hum. Comput. Interact.}} \bibinfo{volume}{1},
  \bibinfo{number}{2} (\bibinfo{year}{1985}), \bibinfo{pages}{133--161}.
\newblock
\urldef\tempurl%
\url{https://doi.org/10.1207/S15327051HCI0102\_3}
\showDOI{\tempurl}


\bibitem[\protect\citeauthoryear{Catolino, Palomba, Zaidman, and
  Ferrucci}{Catolino et~al\mbox{.}}{2019}]%
        {DBLP:journals/jss/CatolinoPZF19}
\bibfield{author}{\bibinfo{person}{Gemma Catolino}, \bibinfo{person}{Fabio
  Palomba}, \bibinfo{person}{Andy Zaidman}, {and} \bibinfo{person}{Filomena
  Ferrucci}.} \bibinfo{year}{2019}\natexlab{}.
\newblock \showarticletitle{Not all bugs are the same: Understanding,
  characterizing, and classifying bug types}.
\newblock \bibinfo{journal}{\emph{J. Syst. Softw.}}  \bibinfo{volume}{152}
  (\bibinfo{year}{2019}), \bibinfo{pages}{165--181}.
\newblock
\urldef\tempurl%
\url{https://doi.org/10.1016/J.JSS.2019.03.002}
\showDOI{\tempurl}


\bibitem[\protect\citeauthoryear{Clegg, Rojas, and Fraser}{Clegg
  et~al\mbox{.}}{2017}]%
        {DBLP:conf/icse/CleggRF17}
\bibfield{author}{\bibinfo{person}{Benjamin~S. Clegg},
  \bibinfo{person}{Jos{\'{e}}~Miguel Rojas}, {and} \bibinfo{person}{Gordon
  Fraser}.} \bibinfo{year}{2017}\natexlab{}.
\newblock \showarticletitle{Teaching Software Testing Concepts Using a Mutation
  Testing Game}. In \bibinfo{booktitle}{\emph{39th {IEEE/ACM} International
  Conference on Software Engineering: Software Engineering Education and
  Training Track, {ICSE-SEET} 2017, Buenos Aires, Argentina, May 20-28, 2017}}.
  \bibinfo{publisher}{{IEEE} Computer Society}, \bibinfo{pages}{33--36}.
\newblock
\urldef\tempurl%
\url{https://doi.org/10.1109/ICSE-SEET.2017.1}
\showDOI{\tempurl}


\bibitem[\protect\citeauthoryear{Connolly, Boyle, MacArthur, Hainey, and
  Boyle}{Connolly et~al\mbox{.}}{2012}]%
        {DBLP:journals/ce/ConnollyBMHB12}
\bibfield{author}{\bibinfo{person}{Thomas~M. Connolly},
  \bibinfo{person}{Elizabeth~A. Boyle}, \bibinfo{person}{Ewan MacArthur},
  \bibinfo{person}{Thomas Hainey}, {and} \bibinfo{person}{James~M. Boyle}.}
  \bibinfo{year}{2012}\natexlab{}.
\newblock \showarticletitle{A systematic literature review of empirical
  evidence on computer games and serious games}.
\newblock \bibinfo{journal}{\emph{Comput. Educ.}} \bibinfo{volume}{59},
  \bibinfo{number}{2} (\bibinfo{year}{2012}), \bibinfo{pages}{661--686}.
\newblock
\urldef\tempurl%
\url{https://doi.org/10.1016/J.COMPEDU.2012.03.004}
\showDOI{\tempurl}


\bibitem[\protect\citeauthoryear{Cooper}{Cooper}{2023}]%
        {DBLP:books/sp/23/Cooper23}
\bibfield{author}{\bibinfo{person}{Kendra M.~L. Cooper}.}
  \bibinfo{year}{2023}\natexlab{}.
\newblock \showarticletitle{Introduction to Software Engineering for Games in
  Serious Contexts}.
\newblock In \bibinfo{booktitle}{\emph{Software Engineering for Games in
  Serious Contexts - Theories, Methods, Tools, and Experiences}},
  \bibfield{editor}{\bibinfo{person}{Kendra M.~L. Cooper} {and}
  \bibinfo{person}{Antonio Bucchiarone}} (Eds.). \bibinfo{publisher}{Springer},
  \bibinfo{pages}{1--16}.
\newblock
\urldef\tempurl%
\url{https://doi.org/10.1007/978-3-031-33338-5\_1}
\showDOI{\tempurl}


\bibitem[\protect\citeauthoryear{Cui and Shi}{Cui and Shi}{2011}]%
        {cui2011based}
\bibfield{author}{\bibinfo{person}{Xiao Cui} {and} \bibinfo{person}{Hao Shi}.}
  \bibinfo{year}{2011}\natexlab{}.
\newblock \showarticletitle{A*-based pathfinding in modern computer games}.
\newblock \bibinfo{journal}{\emph{International Journal of Computer Science and
  Network Security}} \bibinfo{volume}{11}, \bibinfo{number}{1}
  (\bibinfo{year}{2011}), \bibinfo{pages}{125--130}.
\newblock


\bibitem[\protect\citeauthoryear{de~Souza~Santos, de~Magalh{\~{a}}es,
  da~Silva~Correia{-}Neto, da~Silva, Capretz, and Souza}{de~Souza~Santos
  et~al\mbox{.}}{2017}]%
        {DBLP:conf/esem/SantosMCSCS17}
\bibfield{author}{\bibinfo{person}{Ronnie~Edson de Souza~Santos},
  \bibinfo{person}{Cleyton Vanut~Cordeiro de Magalh{\~{a}}es},
  \bibinfo{person}{Jorge da Silva~Correia{-}Neto}, \bibinfo{person}{Fabio
  Queda~Bueno da Silva}, \bibinfo{person}{Luiz~Fernando Capretz}, {and}
  \bibinfo{person}{Rodrigo E.~C. Souza}.} \bibinfo{year}{2017}\natexlab{}.
\newblock \showarticletitle{Would You Like to Motivate Software Testers? Ask
  Them How}. In \bibinfo{booktitle}{\emph{2017 {ACM/IEEE} International
  Symposium on Empirical Software Engineering and Measurement, {ESEM} 2017,
  Toronto, ON, Canada, November 9-10, 2017}},
  \bibfield{editor}{\bibinfo{person}{Ayse Bener}, \bibinfo{person}{Burak
  Turhan}, {and} \bibinfo{person}{Stefan Biffl}} (Eds.).
  \bibinfo{publisher}{{IEEE} Computer Society}, \bibinfo{pages}{95--104}.
\newblock
\urldef\tempurl%
\url{https://doi.org/10.1109/ESEM.2017.16}
\showDOI{\tempurl}


\bibitem[\protect\citeauthoryear{Deterding, Dixon, Khaled, and Nacke}{Deterding
  et~al\mbox{.}}{2011}]%
        {DBLP:conf/mindtrek/DeterdingDKN11}
\bibfield{author}{\bibinfo{person}{Sebastian Deterding}, \bibinfo{person}{Dan
  Dixon}, \bibinfo{person}{Rilla Khaled}, {and} \bibinfo{person}{Lennart~E.
  Nacke}.} \bibinfo{year}{2011}\natexlab{}.
\newblock \showarticletitle{From game design elements to gamefulness: defining
  "gamification"}. In \bibinfo{booktitle}{\emph{Proceedings of the 15th
  International Academic MindTrek Conference: Envisioning Future Media
  Environments, MindTrek 2011, Tampere, Finland, September 28-30, 2011}},
  \bibfield{editor}{\bibinfo{person}{Artur Lugmayr},
  \bibinfo{person}{Helj{\"{a}} Franssila}, \bibinfo{person}{Christian Safran},
  {and} \bibinfo{person}{Imed Hammouda}} (Eds.). \bibinfo{publisher}{{ACM}},
  \bibinfo{pages}{9--15}.
\newblock
\urldef\tempurl%
\url{https://doi.org/10.1145/2181037.2181040}
\showDOI{\tempurl}


\bibitem[\protect\citeauthoryear{Fraser, Gambi, Kreis, and Rojas}{Fraser
  et~al\mbox{.}}{2019}]%
        {DBLP:conf/sigcse/FraserGKR19}
\bibfield{author}{\bibinfo{person}{Gordon Fraser}, \bibinfo{person}{Alessio
  Gambi}, \bibinfo{person}{Marvin Kreis}, {and}
  \bibinfo{person}{Jos{\'{e}}~Miguel Rojas}.} \bibinfo{year}{2019}\natexlab{}.
\newblock \showarticletitle{Gamifying a Software Testing Course with Code
  Defenders}. In \bibinfo{booktitle}{\emph{Proceedings of the 50th {ACM}
  Technical Symposium on Computer Science Education, {SIGCSE} 2019,
  Minneapolis, MN, USA, February 27 - March 02, 2019}},
  \bibfield{editor}{\bibinfo{person}{Elizabeth~K. Hawthorne},
  \bibinfo{person}{Manuel~A. P{\'{e}}rez{-}Qui{\~{n}}ones},
  \bibinfo{person}{Sarah Heckman}, {and} \bibinfo{person}{Jian Zhang}} (Eds.).
  \bibinfo{publisher}{{ACM}}, \bibinfo{pages}{571--577}.
\newblock
\urldef\tempurl%
\url{https://doi.org/10.1145/3287324.3287471}
\showDOI{\tempurl}


\bibitem[\protect\citeauthoryear{Garousi, Rainer, Jr., and Arcuri}{Garousi
  et~al\mbox{.}}{2020a}]%
        {DBLP:journals/jss/GarousiRLA20}
\bibfield{author}{\bibinfo{person}{Vahid Garousi}, \bibinfo{person}{Austen
  Rainer}, \bibinfo{person}{Per~Lauv{\aa}s Jr.}, {and} \bibinfo{person}{Andrea
  Arcuri}.} \bibinfo{year}{2020}\natexlab{a}.
\newblock \showarticletitle{Software-testing education: {A} systematic
  literature mapping}.
\newblock \bibinfo{journal}{\emph{J. Syst. Softw.}}  \bibinfo{volume}{165}
  (\bibinfo{year}{2020}), \bibinfo{pages}{110570}.
\newblock
\urldef\tempurl%
\url{https://doi.org/10.1016/J.JSS.2020.110570}
\showDOI{\tempurl}


\bibitem[\protect\citeauthoryear{Garousi, Rainer, Lauv{\aa}s~Jr, and
  Arcuri}{Garousi et~al\mbox{.}}{2020b}]%
        {garousi2020software}
\bibfield{author}{\bibinfo{person}{Vahid Garousi}, \bibinfo{person}{Austen
  Rainer}, \bibinfo{person}{Per Lauv{\aa}s~Jr}, {and} \bibinfo{person}{Andrea
  Arcuri}.} \bibinfo{year}{2020}\natexlab{b}.
\newblock \showarticletitle{Software-testing education: A systematic literature
  mapping}.
\newblock \bibinfo{journal}{\emph{Journal of Systems and Software}}
  \bibinfo{volume}{165} (\bibinfo{year}{2020}), \bibinfo{pages}{110570}.
\newblock


\bibitem[\protect\citeauthoryear{Hristova, Misra, Rutter, and Mercuri}{Hristova
  et~al\mbox{.}}{2003}]%
        {DBLP:conf/sigcse/HristovaMRM03}
\bibfield{author}{\bibinfo{person}{Maria Hristova}, \bibinfo{person}{Ananya
  Misra}, \bibinfo{person}{Megan Rutter}, {and} \bibinfo{person}{Rebecca
  Mercuri}.} \bibinfo{year}{2003}\natexlab{}.
\newblock \showarticletitle{Identifying and correcting Java programming errors
  for introductory computer science students}. In
  \bibinfo{booktitle}{\emph{Proceedings of the 34th {SIGCSE} Technical
  Symposium on Computer Science Education, {SIGCSE} 2003, Reno, Nevada, USA,
  February 19-23, 2003}}, \bibfield{editor}{\bibinfo{person}{Scott Grissom},
  \bibinfo{person}{Deborah Knox}, \bibinfo{person}{Daniel~T. Joyce}, {and}
  \bibinfo{person}{Wanda~P. Dann}} (Eds.). \bibinfo{publisher}{{ACM}},
  \bibinfo{pages}{153--156}.
\newblock
\urldef\tempurl%
\url{https://doi.org/10.1145/611892.611956}
\showDOI{\tempurl}


\bibitem[\protect\citeauthoryear{Jamil, Arif, Abubakar, and Ahmad}{Jamil
  et~al\mbox{.}}{2016}]%
        {7814898}
\bibfield{author}{\bibinfo{person}{Muhammad~Abid Jamil},
  \bibinfo{person}{Muhammad Arif}, \bibinfo{person}{Normi Sham~Awang Abubakar},
  {and} \bibinfo{person}{Akhlaq Ahmad}.} \bibinfo{year}{2016}\natexlab{}.
\newblock \showarticletitle{Software Testing Techniques: A Literature Review}.
  In \bibinfo{booktitle}{\emph{2016 6th International Conference on Information
  and Communication Technology for The Muslim World (ICT4M)}}.
  \bibinfo{pages}{177--182}.
\newblock
\urldef\tempurl%
\url{https://doi.org/10.1109/ICT4M.2016.045}
\showDOI{\tempurl}


\bibitem[\protect\citeauthoryear{Kappen and Nacke}{Kappen and Nacke}{2013}]%
        {DBLP:conf/gamification/KappenN13}
\bibfield{author}{\bibinfo{person}{Dennis~L. Kappen} {and}
  \bibinfo{person}{Lennart~E. Nacke}.} \bibinfo{year}{2013}\natexlab{}.
\newblock \showarticletitle{The kaleidoscope of effective gamification:
  deconstructing gamification in business applications}. In
  \bibinfo{booktitle}{\emph{Proceedings of the First International Conference
  on Gameful Design, Research, and Applications, Gamification '13, Toronto,
  Ontario, Canada, October 2-4, 2013}},
  \bibfield{editor}{\bibinfo{person}{Lennart~E. Nacke},
  \bibinfo{person}{Kevin~A. Harrigan}, {and} \bibinfo{person}{Neil~C. Randall}}
  (Eds.). \bibinfo{publisher}{{ACM}}, \bibinfo{pages}{119--122}.
\newblock
\urldef\tempurl%
\url{https://doi.org/10.1145/2583008.2583029}
\showDOI{\tempurl}


\bibitem[\protect\citeauthoryear{Lee}{Lee}{2014}]%
        {DBLP:conf/vl/Lee14}
\bibfield{author}{\bibinfo{person}{Michael~Jongseon Lee}.}
  \bibinfo{year}{2014}\natexlab{}.
\newblock \showarticletitle{Gidget: An online debugging game for learning and
  engagement in computing education}. In \bibinfo{booktitle}{\emph{{IEEE}
  Symposium on Visual Languages and Human-Centric Computing, {VL/HCC} 2014,
  Melbourne, VIC, Australia, July 28 - August 1, 2014}},
  \bibfield{editor}{\bibinfo{person}{Scott~D. Fleming}, \bibinfo{person}{Andrew
  Fish}, {and} \bibinfo{person}{Christopher Scaffidi}} (Eds.).
  \bibinfo{publisher}{{IEEE} Computer Society}, \bibinfo{pages}{193--194}.
\newblock
\urldef\tempurl%
\url{https://doi.org/10.1109/VLHCC.2014.6883051}
\showDOI{\tempurl}


\bibitem[\protect\citeauthoryear{Li, Chan, Denny, Luxton{-}Reilly, and
  Tempero}{Li et~al\mbox{.}}{2019}]%
        {DBLP:conf/ace/LiCDLT19}
\bibfield{author}{\bibinfo{person}{Chen Li}, \bibinfo{person}{Emily Chan},
  \bibinfo{person}{Paul Denny}, \bibinfo{person}{Andrew Luxton{-}Reilly}, {and}
  \bibinfo{person}{Ewan~D. Tempero}.} \bibinfo{year}{2019}\natexlab{}.
\newblock \showarticletitle{Towards a Framework for Teaching Debugging}. In
  \bibinfo{booktitle}{\emph{Proceedings of the Twenty-First Australasian
  Computing Education Conference, Sydney, NSW, Australia, January 29-31,
  2019}}, \bibfield{editor}{\bibinfo{person}{Simon} {and}
  \bibinfo{person}{Andrew Luxton{-}Reilly}} (Eds.). \bibinfo{publisher}{{ACM}},
  \bibinfo{pages}{79--86}.
\newblock
\urldef\tempurl%
\url{https://doi.org/10.1145/3286960.3286970}
\showDOI{\tempurl}


\bibitem[\protect\citeauthoryear{Mann and Whitney}{Mann and Whitney}{1947}]%
        {10.1214/aoms/1177730491}
\bibfield{author}{\bibinfo{person}{H.~B. Mann} {and} \bibinfo{person}{D.~R.
  Whitney}.} \bibinfo{year}{1947}\natexlab{}.
\newblock \showarticletitle{{On a Test of Whether one of Two Random Variables
  is Stochastically Larger than the Other}}.
\newblock \bibinfo{journal}{\emph{The Annals of Mathematical Statistics}}
  \bibinfo{volume}{18}, \bibinfo{number}{1} (\bibinfo{year}{1947}),
  \bibinfo{pages}{50 -- 60}.
\newblock
\urldef\tempurl%
\url{https://doi.org/10.1214/aoms/1177730491}
\showDOI{\tempurl}


\bibitem[\protect\citeauthoryear{McCauley, Fitzgerald, Lewandowski, Murphy,
  Simon, Thomas, and Zander}{McCauley et~al\mbox{.}}{2008}]%
        {DBLP:journals/csedu/McCauleyFLMSTZ08}
\bibfield{author}{\bibinfo{person}{Ren{\'{e}}e McCauley}, \bibinfo{person}{Sue
  Fitzgerald}, \bibinfo{person}{Gary Lewandowski}, \bibinfo{person}{Laurie
  Murphy}, \bibinfo{person}{Beth Simon}, \bibinfo{person}{Lynda Thomas}, {and}
  \bibinfo{person}{Carol Zander}.} \bibinfo{year}{2008}\natexlab{}.
\newblock \showarticletitle{Debugging: a review of the literature from an
  educational perspective}.
\newblock \bibinfo{journal}{\emph{Comput. Sci. Educ.}} \bibinfo{volume}{18},
  \bibinfo{number}{2} (\bibinfo{year}{2008}), \bibinfo{pages}{67--92}.
\newblock
\urldef\tempurl%
\url{https://doi.org/10.1080/08993400802114581}
\showDOI{\tempurl}


\bibitem[\protect\citeauthoryear{Michaeli and Romeike}{Michaeli and
  Romeike}{2019}]%
        {DBLP:conf/wipsce/MichaeliR19}
\bibfield{author}{\bibinfo{person}{Tilman Michaeli} {and} \bibinfo{person}{Ralf
  Romeike}.} \bibinfo{year}{2019}\natexlab{}.
\newblock \showarticletitle{Improving Debugging Skills in the Classroom: The
  Effects of Teaching a Systematic Debugging Process}. In
  \bibinfo{booktitle}{\emph{Proceedings of the 14th Workshop in Primary and
  Secondary Computing Education, WiPSCE 2019, Glasgow, Scotland, UK, October
  23-25, 2019}}. \bibinfo{publisher}{{ACM}}, \bibinfo{pages}{15:1--15:7}.
\newblock
\urldef\tempurl%
\url{https://doi.org/10.1145/3361721.3361724}
\showDOI{\tempurl}


\bibitem[\protect\citeauthoryear{Miljanovic and Bradbury}{Miljanovic and
  Bradbury}{2017}]%
        {DBLP:conf/icer/MiljanovicB17}
\bibfield{author}{\bibinfo{person}{Michael~A. Miljanovic} {and}
  \bibinfo{person}{Jeremy~S. Bradbury}.} \bibinfo{year}{2017}\natexlab{}.
\newblock \showarticletitle{RoboBUG: {A} Serious Game for Learning Debugging
  Techniques}. In \bibinfo{booktitle}{\emph{Proceedings of the 2017 {ACM}
  Conference on International Computing Education Research, {ICER} 2017,
  Tacoma, WA, USA, August 18-20, 2017}},
  \bibfield{editor}{\bibinfo{person}{Josh Tenenberg}, \bibinfo{person}{Donald
  Chinn}, \bibinfo{person}{Judy Sheard}, {and} \bibinfo{person}{Lauri Malmi}}
  (Eds.). \bibinfo{publisher}{{ACM}}, \bibinfo{pages}{93--100}.
\newblock
\urldef\tempurl%
\url{https://doi.org/10.1145/3105726.3106173}
\showDOI{\tempurl}


\bibitem[\protect\citeauthoryear{Murphy, Lewandowski, McCauley, Simon, Thomas,
  and Zander}{Murphy et~al\mbox{.}}{2008}]%
        {DBLP:conf/sigcse/MurphyLMSTZ08}
\bibfield{author}{\bibinfo{person}{Laurie Murphy}, \bibinfo{person}{Gary
  Lewandowski}, \bibinfo{person}{Ren{\'{e}}e McCauley}, \bibinfo{person}{Beth
  Simon}, \bibinfo{person}{Lynda Thomas}, {and} \bibinfo{person}{Carol
  Zander}.} \bibinfo{year}{2008}\natexlab{}.
\newblock \showarticletitle{Debugging: the good, the bad, and the quirky -- a
  qualitative analysis of novices' strategies}. In
  \bibinfo{booktitle}{\emph{Proceedings of the 39th {SIGCSE} Technical
  Symposium on Computer Science Education, {SIGCSE} 2008, Portland, OR, USA,
  March 12-15, 2008}}, \bibfield{editor}{\bibinfo{person}{J.~D. Dougherty},
  \bibinfo{person}{Susan~H. Rodger}, \bibinfo{person}{Sue Fitzgerald}, {and}
  \bibinfo{person}{Mark Guzdial}} (Eds.). \bibinfo{publisher}{{ACM}},
  \bibinfo{pages}{163--167}.
\newblock
\urldef\tempurl%
\url{https://doi.org/10.1145/1352135.1352191}
\showDOI{\tempurl}


\bibitem[\protect\citeauthoryear{Nayrolles and Hamou{-}Lhadj}{Nayrolles and
  Hamou{-}Lhadj}{2018}]%
        {DBLP:conf/icse/NayrollesH18}
\bibfield{author}{\bibinfo{person}{Mathieu Nayrolles} {and}
  \bibinfo{person}{Abdelwahab Hamou{-}Lhadj}.} \bibinfo{year}{2018}\natexlab{}.
\newblock \showarticletitle{Towards a classification of bugs to facilitate
  software maintainability tasks}. In \bibinfo{booktitle}{\emph{Proceedings of
  the 1st International Workshop on Software Qualities and Their Dependencies,
  SQUADE@ICSE 2018, Gothenburg, Sweden, May 28, 2018}},
  \bibfield{editor}{\bibinfo{person}{S{\'{e}}verine Sentilles},
  \bibinfo{person}{Barry~W. Boehm}, \bibinfo{person}{Catia Trubiani},
  \bibinfo{person}{Xavier Franch}, {and} \bibinfo{person}{Anne Koziolek}}
  (Eds.). \bibinfo{publisher}{{ACM}}, \bibinfo{pages}{25--32}.
\newblock
\urldef\tempurl%
\url{https://doi.org/10.1145/3194095.3194101}
\showDOI{\tempurl}


\bibitem[\protect\citeauthoryear{Park and Cheon}{Park and Cheon}{2025}]%
        {park2025exploring}
\bibfield{author}{\bibinfo{person}{Eunsung Park} {and} \bibinfo{person}{Jongpil
  Cheon}.} \bibinfo{year}{2025}\natexlab{}.
\newblock \showarticletitle{Exploring Debugging Challenges and Strategies Using
  Structural Topic Model: A Comparative Analysis of High and Low-Performing
  Students}.
\newblock \bibinfo{journal}{\emph{Journal of Educational Computing Research}}
  \bibinfo{volume}{62}, \bibinfo{number}{8} (\bibinfo{year}{2025}),
  \bibinfo{pages}{2104--2126}.
\newblock


\bibitem[\protect\citeauthoryear{Pea}{Pea}{1986}]%
        {pea1986language}
\bibfield{author}{\bibinfo{person}{Roy~D Pea}.}
  \bibinfo{year}{1986}\natexlab{}.
\newblock \showarticletitle{Language-independent conceptual “bugs” in
  novice programming}.
\newblock \bibinfo{journal}{\emph{Journal of educational computing research}}
  \bibinfo{volume}{2}, \bibinfo{number}{1} (\bibinfo{year}{1986}),
  \bibinfo{pages}{25--36}.
\newblock


\bibitem[\protect\citeauthoryear{Quintero and {\'{A}}lvarez}{Quintero and
  {\'{A}}lvarez}{2023}]%
        {DBLP:journals/ieee-rita/QuinteroA23}
\bibfield{author}{\bibinfo{person}{Carlos~Andr{\'{e}}sCaldas Quintero} {and}
  \bibinfo{person}{Gary Alberto~Cifuentes {\'{A}}lvarez}.}
  \bibinfo{year}{2023}\natexlab{}.
\newblock \showarticletitle{Serious Games and Computer Programming Competencies
  Development in Educational Contexts}.
\newblock \bibinfo{journal}{\emph{Rev. Iberoam. de Tecnol. del Aprendiz.}}
  \bibinfo{volume}{18}, \bibinfo{number}{1} (\bibinfo{year}{2023}),
  \bibinfo{pages}{48--53}.
\newblock
\urldef\tempurl%
\url{https://doi.org/10.1109/RITA.2023.3250504}
\showDOI{\tempurl}


\bibitem[\protect\citeauthoryear{Raj and Kumar}{Raj and Kumar}{2022}]%
        {DBLP:journals/inroads/RajK22}
\bibfield{author}{\bibinfo{person}{Rajendra~K. Raj} {and}
  \bibinfo{person}{Amruth~N. Kumar}.} \bibinfo{year}{2022}\natexlab{}.
\newblock \showarticletitle{Toward computer science curricular guidelines 2023
  {(CS2023)}}.
\newblock \bibinfo{journal}{\emph{Inroads}} \bibinfo{volume}{13},
  \bibinfo{number}{4} (\bibinfo{year}{2022}), \bibinfo{pages}{22--25}.
\newblock
\urldef\tempurl%
\url{https://doi.org/10.1145/3571092}
\showDOI{\tempurl}


\bibitem[\protect\citeauthoryear{Reynolds et~al\mbox{.}}{Reynolds
  et~al\mbox{.}}{1999}]%
        {reynolds1999steering}
\bibfield{author}{\bibinfo{person}{Craig~W Reynolds} {et~al\mbox{.}}}
  \bibinfo{year}{1999}\natexlab{}.
\newblock \showarticletitle{Steering behaviors for autonomous characters}. In
  \bibinfo{booktitle}{\emph{Game developers conference}},
  Vol.~\bibinfo{volume}{1999}. Citeseer, \bibinfo{pages}{763--782}.
\newblock


\bibitem[\protect\citeauthoryear{Straubinger, Bloch, and Fraser}{Straubinger
  et~al\mbox{.}}{2024}]%
        {DBLP:conf/icst/StraubingerBF24}
\bibfield{author}{\bibinfo{person}{Philipp Straubinger}, \bibinfo{person}{Lena
  Bloch}, {and} \bibinfo{person}{Gordon Fraser}.}
  \bibinfo{year}{2024}\natexlab{}.
\newblock \showarticletitle{Engaging Young Learners with Testing Using the Code
  Critters Mutation Game}. In \bibinfo{booktitle}{\emph{{IEEE} International
  Conference on Software Testing, Verification and Validation, {ICST} 2024 -
  Workshops, Toronto, ON, Canada, May 27-31, 2024}}.
  \bibinfo{publisher}{{IEEE}}, \bibinfo{pages}{322--330}.
\newblock
\urldef\tempurl%
\url{https://doi.org/10.1109/ICSTW60967.2024.00063}
\showDOI{\tempurl}


\bibitem[\protect\citeauthoryear{Straubinger, Caspari, and Fraser}{Straubinger
  et~al\mbox{.}}{2023}]%
        {DBLP:conf/icst/StraubingerCF23}
\bibfield{author}{\bibinfo{person}{Philipp Straubinger}, \bibinfo{person}{Laura
  Caspari}, {and} \bibinfo{person}{Gordon Fraser}.}
  \bibinfo{year}{2023}\natexlab{}.
\newblock \showarticletitle{Code Critters: {A} Block-Based Testing Game}. In
  \bibinfo{booktitle}{\emph{{IEEE} International Conference on Software
  Testing, Verification and Validation, {ICST} 2023 - Workshops, Dublin,
  Ireland, April 16-20, 2023}}. \bibinfo{publisher}{{IEEE}},
  \bibinfo{pages}{426--429}.
\newblock
\urldef\tempurl%
\url{https://doi.org/10.1109/ICSTW58534.2023.00077}
\showDOI{\tempurl}


\bibitem[\protect\citeauthoryear{Straubinger and Fraser}{Straubinger and
  Fraser}{2023}]%
        {DBLP:conf/issre/StraubingerF23}
\bibfield{author}{\bibinfo{person}{Philipp Straubinger} {and}
  \bibinfo{person}{Gordon Fraser}.} \bibinfo{year}{2023}\natexlab{}.
\newblock \showarticletitle{A Survey on What Developers Think About Testing}.
  In \bibinfo{booktitle}{\emph{34th {IEEE} International Symposium on Software
  Reliability Engineering, {ISSRE} 2023, Florence, Italy, October 9-12, 2023}}.
  \bibinfo{publisher}{{IEEE}}, \bibinfo{pages}{80--90}.
\newblock
\urldef\tempurl%
\url{https://doi.org/10.1109/ISSRE59848.2023.00075}
\showDOI{\tempurl}


\bibitem[\protect\citeauthoryear{Toda, Valle, and Isotani}{Toda
  et~al\mbox{.}}{2017}]%
        {DBLP:conf/hefa/TodaVI17}
\bibfield{author}{\bibinfo{person}{Armando~Maciel Toda}, \bibinfo{person}{Pedro
  Henrique~Dias Valle}, {and} \bibinfo{person}{Seiji Isotani}.}
  \bibinfo{year}{2017}\natexlab{}.
\newblock \showarticletitle{The Dark Side of Gamification: An Overview of
  Negative Effects of Gamification in Education}. In
  \bibinfo{booktitle}{\emph{Higher Education for All. From Challenges to Novel
  Technology-Enhanced Solutions - First International Workshop on Social,
  Semantic, Adaptive and Gamification Techniques and Technologies for Distance
  Learning, {HEFA} 2017, Macei{\'{o}}, Brazil, March 20-24, 2017, Revised
  Selected Papers}} \emph{(\bibinfo{series}{Communications in Computer and
  Information Science}, Vol.~\bibinfo{volume}{832})},
  \bibfield{editor}{\bibinfo{person}{Alexandra~Ioana Cristea},
  \bibinfo{person}{Ig~Ibert Bittencourt}, {and} \bibinfo{person}{Fernanda
  Lima}} (Eds.). \bibinfo{publisher}{Springer}, \bibinfo{pages}{143--156}.
\newblock
\urldef\tempurl%
\url{https://doi.org/10.1007/978-3-319-97934-2\_9}
\showDOI{\tempurl}


\bibitem[\protect\citeauthoryear{Valle, Toda, Barbosa, and Maldonado}{Valle
  et~al\mbox{.}}{2017}]%
        {DBLP:conf/fie/ValleTBM17}
\bibfield{author}{\bibinfo{person}{Pedro Henrique~Dias Valle},
  \bibinfo{person}{Armando~Maciel Toda}, \bibinfo{person}{Ellen~Francine
  Barbosa}, {and} \bibinfo{person}{Jos{\'{e}}~Carlos Maldonado}.}
  \bibinfo{year}{2017}\natexlab{}.
\newblock \showarticletitle{Educational games: {A} contribution to software
  testing education}. In \bibinfo{booktitle}{\emph{2017 {IEEE} Frontiers in
  Education Conference, {FIE} 2017, Indianapolis, IN, USA, October 18-21,
  2017}}. \bibinfo{publisher}{{IEEE} Computer Society}, \bibinfo{pages}{1--8}.
\newblock
\urldef\tempurl%
\url{https://doi.org/10.1109/FIE.2017.8190470}
\showDOI{\tempurl}


\bibitem[\protect\citeauthoryear{Yamoul, Ouchaouka, Moussetad, and
  Radid}{Yamoul et~al\mbox{.}}{2023}]%
        {DBLP:conf/cist/YamoulOMR23}
\bibfield{author}{\bibinfo{person}{Soumia Yamoul}, \bibinfo{person}{Lynda
  Ouchaouka}, \bibinfo{person}{Mohammed Moussetad}, {and}
  \bibinfo{person}{Mohamed Radid}.} \bibinfo{year}{2023}\natexlab{}.
\newblock \showarticletitle{Systematic Review of Serious Games in Higher
  Education: Objectives, Benefits, Limitations, and Perspectives}. In
  \bibinfo{booktitle}{\emph{7th {IEEE} Congress on Information Science and
  Technology, CiSt 2023, Agadir - Essaouira, Morocco, December 16-22, 2023}}.
  \bibinfo{publisher}{{IEEE}}, \bibinfo{pages}{450--455}.
\newblock
\urldef\tempurl%
\url{https://doi.org/10.1109/CIST56084.2023.10409880}
\showDOI{\tempurl}


\end{thebibliography}
	
\end{document}